\documentclass[sigconf]{acmart}
\renewcommand\footnotetextcopyrightpermission[1]{} 
\setcopyright{none}
\usepackage{array,ragged2e}

\usepackage{color}
\usepackage{xspace}
\usepackage{url}

\newif\ifsubmit
\submitfalse

\ifsubmit
    \newcommand{\amir}[1]{}
     
\else
    \newcommand{\amir}[1]{\textcolor{blue}{Amir: #1}}
    
\fi

\usepackage{array}
\newcommand{\PreserveBackslash}[1]{\let\temp=\\#1\let\\=\temp}
\newcolumntype{C}[1]{>{\PreserveBackslash\centering}p{#1}}
\newcolumntype{R}[1]{>{\PreserveBackslash\raggedleft}p{#1}}
\newcolumntype{L}[1]{>{\PreserveBackslash\raggedright}p{#1}}

\usepackage{algorithm}
\usepackage[normalem]{ulem}
\usepackage{verbatim}
\usepackage{url}
\usepackage{enumitem}
\usepackage{tikz}
\usepackage{caption}
\usepackage{subcaption}
\usepackage{xspace}
\usepackage{csquotes}

\usepackage{tikz}

\def\BibTeX{{\rm B\kern-.05em{\sc i\kern-.025em b}\kern-.08emT\kern-.1667em\lower.7ex\hbox{E}\kern-.125emX}}

\usepackage[noend]{algpseudocode}

\makeatletter
\def\BState{\State\hskip-\ALG@thistlm}
\makeatother

\newcommand{\blue}[1]{\textcolor{black}{#1}}

\newcommand{\soteria}[0]{\texttt{Soteria}\xspace}
\newcommand{\soterianospace}[0]{\soteria}

\newcommand{\secintentnospace}[0]{\secintent}
\newcommand{\secintent}[0]{\texttt{VISCR}\xspace}

\newcommand{\vicenospace}[0]{\vice}
\newcommand{\vice}[0]{\texttt{VICE}\xspace}

\newcommand{\ballnumber}[1]{\tikz[baseline=(myanchor.base)] \node[circle,fill=.,inner sep=1pt] (myanchor) {\color{-.}\bfseries\footnotesize #1};}

\newcommand{\numofapps}{907\xspace}
\newcommand{\soteriaruntime}{$\sim$1147\xspace}

\newenvironment{packeditemize}{
\begin{list}{$\bullet$}{
\setlength{\itemsep}{1.5pt}
\setlength{\labelwidth}{8pt}
\setlength{\leftmargin}{10pt}
\setlength{\labelsep}{3pt}
\setlength{\listparindent}{\parindent}
\setlength{\parsep}{1.5pt}
\setlength{\parskip}{1.5pt}
\setlength{\topsep}{1.5pt}}}{\end{list}}

\title{ VISCR: Intuitive \& Conflict-free Automation for Securing the Dynamic Consumer IoT Infrastructures}

\begin{document}\sloppy

\author{Vasudevan Nagendra}
\affiliation{\institution{Stony Brook University}}
\email{vnagendra@cs.stonybrook.edu}

\author{Arani Bhattacharya}
\affiliation{\institution{Stony Brook University}}
\email{arbhattachar@cs.stonybrook.edu}

\author{Vinod Yegneswaran}
\affiliation{\institution{SRI International}}
\email{vinod@csl.sri.com}

\author{Amir Rahmati}
\affiliation{\institution{Stony Brook University}}
\email{amir@cs.stonybrook.edu}


\author{Samir R Das}
\affiliation{\institution{Stony Brook University}}
\email{samir@cs.stonybrook.edu}

\thispagestyle{plain}
\pagestyle{plain}

\begin{abstract}
Consumer IoT is characterized by heterogeneous devices with diverse functionality and programming interfaces. This lack of homogeneity makes the integration and security management of IoT infrastructures a daunting task for users and administrators.  In this paper, we introduce \secintent, a Vendor-Independent policy Specification and Conflict Resolution engine that enables {\it conflict-free policy specification and enforcement} in IoT environments.
\secintent converts the topology of the IoT infrastructure into a tree-based abstraction and translates existing policies from heterogeneous vendor-specific programming languages such as Groovy-based SmartThings, OpenHAB, IFTTT-based templates, and MUD-based profiles into a vendor-independent graph-based specification. Using the two,
\secintent can automatically detect rouge policies, conflicts, and bugs for coherent automation. Upon detection, \secintent infers new policies and proposes them to users as alternatives to existing policies for fine-tuning and conflict-free enforcement. 
We evaluated \secintent using a dataset of \numofapps IoT apps, programmed using heterogeneous automation specifications in a simulated smart-building IoT infrastructure. In our experiments, among \numofapps IoT apps, \secintent exposed 342 of IoT apps as exhibiting one or more violations. \secintent detected 100\% of violations reported by existing state-of-the-art tool, while detecting new types of violations in an additional 266 apps.
In terms of performance, \secintent can generate 400 abstraction trees (used in specifying policies) with 100K leaf nodes in $<$1.2sec. In our experiments, \secintent took 80.7 seconds to analyze our infrastructure of \numofapps apps; a 14.2$\times$ reduction compared to the state-of-the-art. After the initial analysis, \secintent is capable of adopting new policies in sub-second latency to handle changes.

\end{abstract}

\maketitle

\section{Introduction}\label{sec:intro}
The adoption of the Internet of Things (IoT) has led to an explosion in the number of devices integrated into consumer IoT infrastructures (e.g., ``smart homes'', ``smart campus'', and ``smart cities'') ~\cite{gartner-future-smart-home,gartner_2020}. The heterogeneity of these infrastructures poses two major challenges in developing and enforcing policies across them: 
\begin{packeditemize}
\item [$(1)$]
{\it Coherent Policy Expression}:  Today, IoT device vendors support web and mobile-based apps, wide range of IoT automation frameworks, and specification languages~\cite{openhab,apple_iot_homekit,groovy_smartthingsapps,mud_spec,decouple_ifttt} that allow users and administrators to program their IoT infrastructures. However, directly capturing the high-level automation (or policy) intent of IoT administrators, using these vendor-specific IoT apps, is a challenging task as it requires administrators of the IoT infrastructure to manually decompose their high-level intents into device-specific rules prior to installation onto IoT infrastructures. 

\item [$(2)$]
{\it Conflict-free Enforcement}: Consumer IoT infrastructures are programmed by multiple administrators having complex {\it roles} and varying levels of {\it skill}, which may include novice smart-home users (e.g., parents, kids), smart-campus or smart-city administrators (e.g., HVAC admins, fire-safety admins). Configuring {\it conflict-free} automation in such multi-administrative environments is a tedious process. 
\end{packeditemize}

\vspace{2pt}
Recent studies independently underscore the need for new access control and policy frameworks to address specifically the security goals of IoT ecosystem~\cite{rethink_acls_iot,soteria,celikiotguard}. These solutions are designed to identify violations in IoT device ecosystems with {\it homogeneous} programming specifications, which does not apply to contemporary consumer IoT deployments that commonly involve devices using diverse vendor-specific APIs and heterogeneous programming specifications. Furthermore, existing tools either rely on model checking for static analysis of IoT automation programs to detect the conflicts~\cite{soteria}, or require IoT automation code to be instrumented for detecting run-time violations~\cite{celikiotguard}. Existing tools and techniques also leave a wide spectrum of automation bugs and conflicts undetected. For example, {\it gap in automation} due to lack of expertise among administrators, {\it rogue policies} on infrastructures that the user does not control, violations that might arise due to loops among automation rules, and other potential run-time violations are some of the key issues left unaddressed by the existing automation frameworks (discussed in \S \ref{sec:composition_security_analysis}).

In this paper, we introduce ``\secintentnospace'', a new intent-based IoT policy automation framework.
For effectively accommodating the automation rules and policies that are specified in the existing IoT infrastructure using heterogeneous programming specification languages, \secintent builds vendor-independent graph-based specification model. \secintent translates the IoT automation programs specified using groovy-based programs for smarthings~\cite{groovy_smartthingsapps}, OpenHAB-based rules~\cite{openhab}, MUD-profiles~\cite{mud_spec} and IFTTT-based applets~\cite{decouple_ifttt} into vendor-independent graph-based specifications (\S~\ref{sec:vi_spec}). In addition, \secintent allows administrators of an IoT infrastructure to directly and succinctly capture their dynamic automation use-cases and policy intents in a vendor-independent manner using simple and intuitive graph-based abstractions and specification mechanism (\S~\ref{sec:abstractions_graph-based-specification}).
\secintent uses these graphs to detect bugs and conflicts across policies that could otherwise go undetected using existing model checking-based tools~\cite{soteria,celikiotguard}. These include: ($a$) static compile-time conflicts, ($b$) gap in the automation, ($c$) conflicts due to loops among automation rules, ($d$) code sanity bugs, ($e$) access violations, and ($f$) rogue policies. 

\secintent supports {\tt PRECEDENCE} operator, for automatically resolving the conflicts,  while unresolved conflicts are forwarded to appropriate IoT users or administrators for manual resolution. The policy reconciliation engine decomposes the composed conflict-free policy graph into set of device-specific rules for enforcement onto actual devices (e.g., IoT devices, IoT Gateways) as IoT apps and ACL rules. In addition, \secintent infers new policy and propose it to IoT users for fine-tuning the existing policies, while addressing the security and safety violations that arise among the policies (\S \ref{sec:policy_enforcement}).
We evaluate \secintent on a simulated smart-building IoT infrastructure with \numofapps apps. \secintent detected a wide range of bugs i.e., $\approx$37.7\% of IoT apps are reported for one or more conflicts and bugs compared to $\approx$8.4\% of static compile-time conflicts detected with existing model checking-based tool~\cite{soteria} while incurring less than 3.8\% false positives. We discuss the resultant conflicts and bugs detected by \secintent and their categories in \S \ref{sec:prototype_evaluations}.

\noindent In summary, our paper makes the following key contributions:

\begin{packeditemize}

\item
We design \secintent, a vendor-independent  graph-based policy specification mechanism that translates automation rules and policies specified using heterogeneous programming specification languages into vendor-independent graph-based policies. We carry out security analysis of \secintent to evaluate the {\it code sanity} and detect bugs present in the code. We develop intuitive policy specification mechanism using tree-based abstractions that allows users to directly capture the vendor-independent graph-based policies for automation and policy specification (\S \ref{sec:vi_spec_code_analysis}).

\item
\secintent detects bugs and conflicts that arise among the policies represented in vendor-independent graph-based specification and resolve them using {\tt PRECEDENCE} operation for conflict-free enforcement. We developed techniques to automatically infer new policies for proposing it to user for fine-tuning the existing rules (\S \ref{sec:composition_security_analysis}).

\item
We evaluate \secintent using \numofapps IoT apps (i.e., both vetted and unvetted) in a simulated smart building infrastructure with real world automation use cases reported by IoT users and administrators in their IoT infrastructures and compare with existing techniques (\S \ref{sec:prototype_evaluations}).

\end{packeditemize}


\section{Background and Motivation}\label{sec:background_motivation}

\label{sec:background}

A wide variety of automation frameworks and specification languages are supported by vendors for automating IoT infrastructures~\cite{groovy_smartthingsapps,openhab,apple_iot_homekit,decouple_ifttt,mud_spec}. 
Such heterogeneity makes programming the IoT infrastructure a challenging task.

In addition, the presence of multiple administrators\footnote{For example, the smart home automation rules and policies are specified by the members of the family (e.g., parents, kids, guests), while smart campus and enterprise IoT infrastructures are managed by different types of IoT administrators such as HVAC, fire-safety, utilities \& energy, and infrastructure (or building) administrators.} in IoT infrastructures with different roles \& responsibilities further complicates the situation resulting in conflicts and errors among IoT infrastructure automation. Hence, the {\it heterogeneity}, and lack of a {\it unified policy-specification} in multi-administrative IoT infrastructures forces administrators to independently develop automation policies and 
exchange their policies offline. Such independently-developed policies, which are often exchanged offline are manually inspected for detecting the conflicts and violations. For effectively automating the IoT infrastructures and governing the communication among its devices following policies are required: 


\vspace{3pt}
\noindent \textbf{Trigger- \& Action-based Rules}: Automation rules that allow the administrator to specify policies on a set of devices that perform some action in response to an event (e.g., ). Example of programming specification mechanisms used for these type of policies includes SmartThings Groovy-based rules~\cite{groovy_smartthingsapps}, OpenHAB-based rules~\cite{openhab}, IFTTT-based applets~\cite{decouple_ifttt}, and Apple HomeKit~\cite{apple_iot_homekit_swift}.

\vspace{3pt}
\noindent \textbf{ACL-based Policies}:  ACL-based policies restrict the type of communication the IoT devices are permitted, during normal and abnormal states, within the IoT ecosystem. Examples include MUD-based profiles~\cite{mud_spec} and traditional IP-based filtering rules. 



\begin{figure}[tbp!]
\centering
\includegraphics{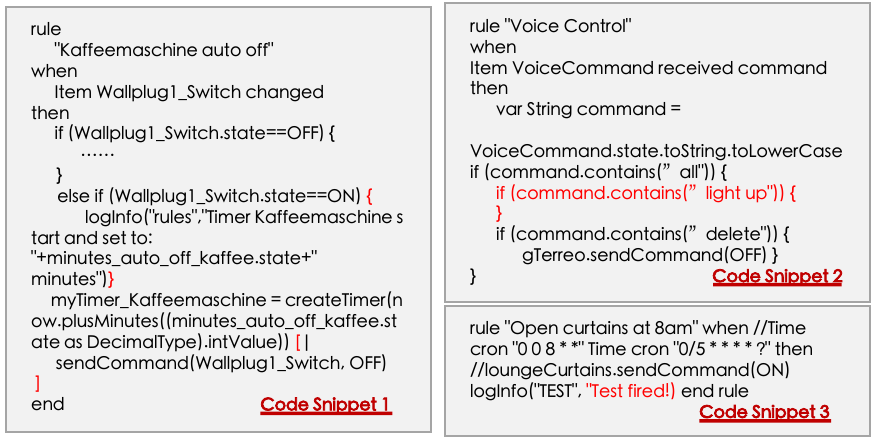}
\vspace{-20pt}
\caption{\small IoT device misbehavior due to commonly seen code sanity issues~\cite{openhab_missing_quotes,openhab_missing_if_braces,openhab_misbehaving_val_var_references}: (1) Code snippet 1: Misaligned parenthesis leading to log incorrectly reporting the state change of coffee machine
(2) Code snippet 2: Illustrating empty If/Else blocks, (3) Code snippet 3: Missing conditional block closures or missing quotes. Some code might hidden in the above examples for brevity. 
}\label{figure:code_sanity_undefined_unfererenced}
\end{figure}

\subsection{Intuitive  Policy Specification}\label{sec:motive_policy_specification} 

IoT infrastructures are typically managed by multiple administrators, each of them responsible for the management of a specific group of devices. For effectively capturing policy intents from multiple administrators, a policy framework should support following capabilities: ($i$) Ability to logically group devices in accordance with the policy requirements of each of the administrator. For example, the IoT administrators handling cameras of BLDG1 and BLDG2 of the campus network should have abstractions that logically group the devices belonging to those locations; ($ii$) Provide isolation among administrators while exposing only the necessary abstractions required for policy specification avoiding {\it rogue policies} from being specified. For example, a fire-safety administrator should not be exposed to other IoT infrastructure details (e.g., cameras, HVACs etc.) unless cross-device policy specification is required. 

Currently, it is challenging to provide logical isolation across different policy administrators and the devices they administer, which motivates the need for fine-grained abstractions and logical grouping of IoT devices as shown in Figure~\ref{fig:sample_network_abstractions}. Similarly, as shown in Figure \ref{figure:code_sanity_undefined_unfererenced}, using heterogeneous automation specification languages 
in the IoT infrastructures is prone to errors due to administrator's lack of expertise in programming~\cite{openhab_missing_quotes,openhab_missing_if_braces,smartthings_errors_community,ifttt_errors_community,applehomekit_errors_community}. 


\vspace{2pt}
\noindent $\blacksquare$ \textbf{Research Goal:} The aforementioned challenges motivate the need for an {\it intuitive graph-based specification} mechanism that is {\it vendor-independent} and allows IoT administrators to easily capture their policy intents while avoiding the steep learning curves with existing automation specification languages.

\subsection{Conflict Detection \& Resolution}\label{motive:conflict_detect_resolve}

Policy intents of an IoT administrator are implemented using discrete automation rules configured onto each device using the mobile or web-based user interfaces, and automation specification languages~\cite{groovy_smartthingsapps,openhab,mud_spec,samsung_ifttt}. The existing consumer IoT ecosystem lacks means to effectively detect the conflicts, bugs and violations that arise among the policies (i.e., captured independently by each of the administrators using heterogeneous specification languages) and reconcile them onto each of the IoT devices.

\begin{table}[tbp!]
  \footnotesize
  \centering
\begin{tabular}{|p{0.06in}|p{0.27in}|p{0.30in}|p{2.1in}|} \hline
& \textbf{Eco-system} & \textbf{Goal} & \textbf{Policy Intents of the IoT Administrator}  \\ \hline
P1 & Campus & Safety, Privacy & Allow video feeds from camera to be sent to fire-safety admins in  event of fire alarm. \\ \hline
P2 & Campus & Security & Cameras for active monitoring is turned OFF between 9 PM and 7AM. Cameras are turned ON only when motion is detected. \\ \hline
\multicolumn{4}{|p{3.2in}|}{{\em \textbf{Conflict}:  Policies P1 and P2 leads to potential security \& safety violation in smart campus. When video feed is shared with fire-safety staff, the video feed is interrupted by the camera's idle-time event which turns OFF the camera.}}\\ \hline
P3 & Home & Safety & Between 10 PM and 7 AM open the main door to kids and guests only with authorized user's approval (e.g., parents). 
\\ \hline
P4 & Home & Safety & In case of a fire-alarm event, initially warn users, and then open the windows and doors to allow residents to exit safely.  \\ \hline
\multicolumn{4}{|p{3.2in}|}{{\em \textbf{Conflict}:  Concurrent activation of policies P3 and P4 in a smart home leads to a potential safety violation. In case of a fire alarm after 10 P.M.,  the IoT automation keeps the door locked due to policy P3, preventing kids from leaving and fire-safety officers from entering.}}\\ \hline
	\end{tabular}
\caption{\small Policy conflict examples in consumer IoT Infrastructures.}\label{table:policy-intents}
\vspace{-15pt}
\end{table}

Recent studies have demonstrated that major vulnerabilities in the IoT infrastructure are commonly due to simple human errors and lack of expertise in policy configuration~\cite{iot_vulnerability1,iot_vulnerability2,openhab_missing_quotes,openhab_missing_if_braces}.
As highlighted in Table \ref{table:policy-intents} \& Table \ref{table:smart-building-polcy-conflicts}, there are range of automation bugs, policy conflicts and violations (e.g., gap in automation, potential run-time conflicts, and code sanity bugs) that could go undetected with existing conflict detection techniques~\cite{celikiotguard,soteria,pga_sigcomm15}. Let us consider following policy conflicts: 


\vspace{2pt}
\noindent {\em Policy Conflict 1}: Consider for example two policies P3 and P4 with conflicting actions in opening main door (Table \ref{table:policy-intents}). In case of a fire alarm after 10 P.M., (both policies P3 and P4 are activated) the current IoT automation keeps the door locked due to policy P3, preventing users from leaving and fire-safety officers from entering. Proactively detecting such conflicting actions resulting in safety violations in compile-time is a challenging task.

\vspace{3pt}
\noindent {\em Policy Conflict 2}: As shown in Table \ref{table:policy-intents}, policies P1 and P2 result in run-time violation. When video feed is shared with fire-safety staff, the video feed is interrupted by the camera's idle-time event which turns OFF the camera interrupting the video feed from being shared, leading to safety violations. A major concern with such policies is that they are activated in response to an event received from one or more sensors. Such violations are difficult to be proactively detected and resolved in compile-time as the asynchronous inactivity counter is triggered, which depends on environmental conditions, (i.e., time and motion sensing). 



\vspace{2pt}
\noindent $\blacksquare$ \textbf{Research Goal:} The aforementioned examples demonstrate the need for collaboration across administrators of IoT infrastructures for charting out their policies and manually resolving conflicts that arise among their policies before enforcing them onto the IoT infrastructure. Thus, a fundamental requirement for any IoT-based policy framework is to effectively handle the dynamic characteristics of IoT infrastructure, automatically adapt to such dynamic changes and enforce a new set of policies by proactively detecting and resolving conflicts that might arise at run-time.




\section{\secintent System Design}\label{sec:overview_usage_threat_model}
To address the limitations of existing IoT automation and policy framework, we make case for {\it unified intent-based} policy framework that provides vendor-independent policy specification and conflict-free policy enforcement mechanisms for IoT infrastructures.

\subsection{Overview}\label{sec:oview}
\begin{figure}[tbp!]
\centering
\includegraphics[width=1.0\linewidth]{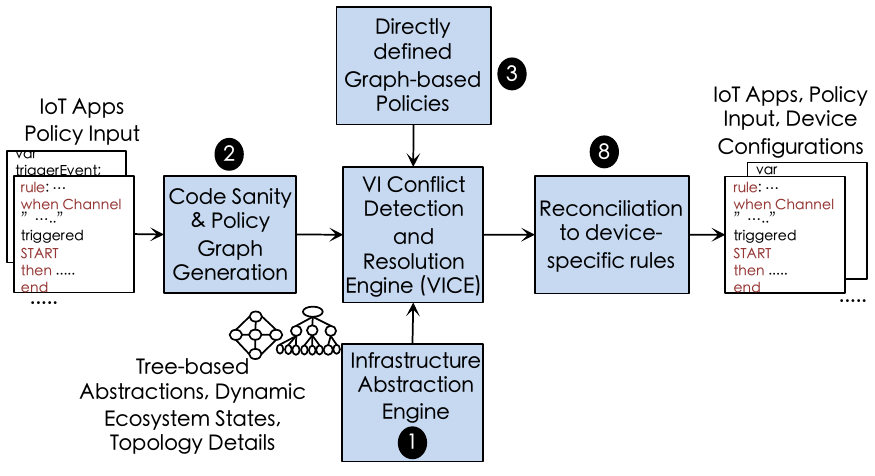}
\vspace{-15pt}
\caption{\secintentnospace: Overall system architecture.}\label{figure:viscr_system_arch}
\vspace{-5pt}
\end{figure}


As shown in Figure \ref{figure:viscr_system_arch}, the automation rules captured using various vendor-specific automation and policy specification languages~\cite{groovy_smartthingsapps,decouple_ifttt,openhab,mud_spec} are provided as input to the code sanity and graph generation module (\ballnumber{2}). As a first step, \secintent analyses IoT automation programs to detect sanity bugs present in it (Section \ref{sec:code_sanity_analysis}). After the initial sanity analysis is performed, \secintent translates the IoT programs (i.e., specified using heterogeneous specification languages) into vendor-independent graph-based specifications. \secintent maintains mapping between the IoT programs, associated device configurations, and graph-based policies as {\it policy mappings}, which is used in policy enforcement i.e., for reconciling composed policy graph to device-specific rules, as discussed in Section \ref{sec:policy_enforcement}. 

For translation of vendor-specific automation rules to vendor-independent graph-based policies we have built necessary lexing and parsing grammar (.g4), and mapping (.map) files specific to each of the automation specification language with ANTLR (\S \ref{sec:vi_spec}). 

To enhance usability of our policy framework, \secintent also supports a graph-based policy specification user interface (\ballnumber{3}) i.e., with simple drag-and-drop option to input policy entities using the tree-based infrastructure abstractions supplied to each of the administrator. This allows IoT users and administrators to easily and intuitively capture their policy intents in a vendor-independent manner without any need to learn multiple IoT automation programming languages. The IoT infrastructure details (i.e., tree-based IoT infrastructure abstractions) necessary for specifying the policies are automatically derived from the existing IoT infrastructure and its cloud data sources (\ballnumber{1}). 

\begin{figure}[tbp!]
\centering
\includegraphics[width=1.0\linewidth]{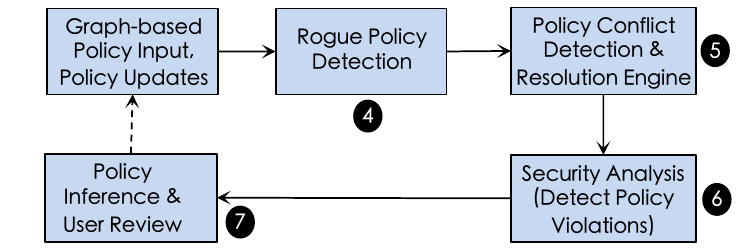}
\vspace{-15pt}
\caption{\vicenospace: Functional flow architecture overview.}\label{figure:vice_arch}
\end{figure}

In the next step, these vendor-independent graph-based policies (from \ballnumber{2} \& \ballnumber{3}) are supplied as input to the vendor-independent conflict detection and resolution engine (\vicenospace) for detecting conflicts, violations and bugs among them these policies. Especially, \vice detects {\it rogue policies} (\ballnumber{4}) that are configured by policy administrators who are not authorized to specify automation rules or policies on any specific portion of the IoT infrastructure (Section \ref{sec:graph-based-specification}). Further, in the next step the policies that are detected for rogue policies are provided as input the policy composition engine (i.e., conflict detection and resolution engine \ballnumber{5}), as shown in Figure \ref{figure:vice_arch}. Policy composition engine uses {\it precedence} mechanism for automatically resolving the conflicts among the policies. Unresolved conflicts and bugs are reported to the administrators of IoT infrastructure for manual resolution. 

The composition engine send the composed policy graph to the security analysis module for further inspection of violations (\ballnumber{6}), which detects following types of conflicts and bugs among the policies that makes infrastructure vulnerable and prone to errors (Section \ref{sec:security_analysis}): ($i$) {\it gap in the automation} that could make infrastructure vulnerable, ($ii$) {\it loops} that exist among automation rules, ($iii$) access violations, and ($iv$) {\it potential conflicts} that could arise in run-time. 

The \secintentnospace's policy inference engine (\ballnumber{7}) automatically infers policies specific to each of the conflicts and propose it to users  for fine-tuning the existing policies (Section \ref{sec:policy_enforcement}). The policies that are fine-tuned are once again composed for detecting the conflicts (\ballnumber{4}). Finally, the identified composed conflict-free policy graphs, and policy mappings are sent to the {\it policy reconciliation module} (\ballnumber{8}) for enforcing the policies as device-specific rules. Finally, the vendor-specific configurations and reconciled IoT apps are generated as outcome of the reconciliation module for installing conflict-free rules on to the IoT infrastructure.

\subsection{Usage Scenario \& Threat Model}\label{sec:usage_threat_model}

\secintentnospace's intended use is within consumer IoT infrastructures, which includes smart home, smart campus, smart city IoT infrastructures, and enterprise networks. Though, the \secintentnospace's architecture is generic enough to accommodate Industrial IoT, but it is out of scope considering different use case scenarios and requirements in industrial ecosystem. 
\secintent aims to proactively reduce the conflicts among policies, violations and bugs that arise in automating the IoT infrastructures. \secintent provides necessary isolation across users or administrators by delegating their roles and responsibilities through supply of dedicated Infrastructure abstraction tress, through which the policies could be specified. In case of conflicts that arise among the policies of different administrators the policies are enforced only after the resolution is made and according to the precedence of its administrators. \secintentnospace's policy inference mechanism  proposes new policies that allows users or administrators to fine-tune their IoT ecosystem or resolve the bugs that exist among its policies. The firmware bugs and vulnerabilities are out of scope of this work.  

Under this usage scenario, we adopt simple and realistic threat model. We trust the IoT devices and its firmware. Hence, we assume that the IoT administrators or users are either novice or could be malicious. We expect skilled administrators can craft policies that could make network vulnerable. Similarly, the novice user's lack of skills could result in IoT infrastructure automation that might result in gap in the automation, which could make network vulnerable. Our threat model considers the conflicts that arise in IoT infrastructures programmed using following programming specification languages \blue{such as IFTTT-based Applets, Groovy-based SmartThings, OpenHAB programs and MUD-profiles}. Other programming specifications such as Apple HomeKit and other programming specifications are out of scope and will be supported by the \secintent framework as part of its future work.

In our threat model, we assume that a subset of IoT users and administrators who program the IoT ecosystem could be malicious.  We attempt to address issues that arise from flaws in the implementation of IoT programs as well as vulnerabilities introduced by malicious users. \secintent proposes to use {\it pessimistic} approach by which administrators can program for safety policies from specific users and for specific set of devices and their capabilities are given higher precedence compared to security and privacy policies of others. Though the precedence is completely programmable, but, it solely depends on users ability to correctly use it, which could at times mask the conflicts. Hence, {\it precedence} operator need to be used diligently by administrators for auto-resolving the conflicts that are detected by \secintentnospace, which could otherwise be safely resolved manually by users.

\section{VI Specification \& IoT App Analysis}\label{sec:vi_spec_code_analysis}
\begin{figure}[tbp!]
\includegraphics{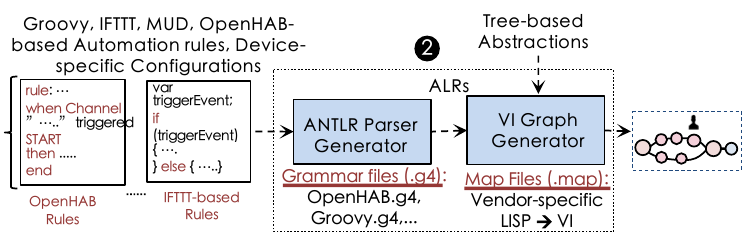}
\vspace{-15pt}
\caption{\small Functional block diagram of Code sanity and Graph generation Module. Architecture for generating Vendor-Independent graph-based specifications from vendor-specific automation rules.}\label{figure:vise_architecture}
\end{figure}

Realizing coherent automation in existing IoT infrastructures that uses heterogeneous specification languages (i.e., multiple automation frameworks) is a challenging task, which requires administrators to analyze each of the IoT Apps for detecting the conflicts and bugs for manually resolving them. The whole procedure of manually detecting conflicts and resolving them is a tedious process and could be prone to error. 
Hence, we propose a vendor-independent specification engine (\secintent) that serves following key purpose: ($i$) Translates IoT automation programs and policies specified using Groovy, IFTTT, MUD, and OpenHAB-based programs into a vendor-independent graph-based specification, and ($ii$) Analyzes the IoT Apps for its sanity.

\subsection{Vendor-Independent Model}\label{sec:vi_spec}
\begin{figure}[tbp!]
\includegraphics{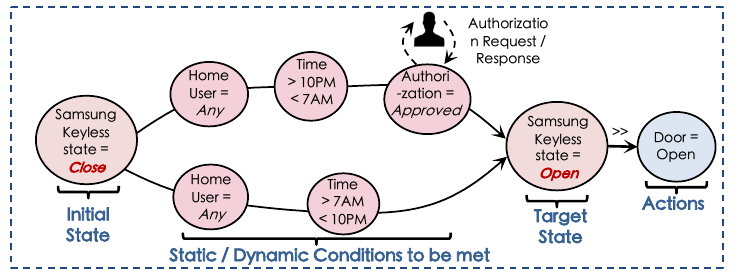}
\vspace{-10pt}
\caption{\small Vendor-independent representation outcome of \secintent for Policy P2 specified using Groovy-based SmartThings app (Table \ref{table:policy-intents}).}\label{fig_vise_translation_policy_p2}
\end{figure}

As illustrated in Figure \ref{figure:viscr_system_arch}, Code sanity and Graph generator module of \secintent engine consumes vendor-specific IoT Apps (i.e., IFTTT-based Applets, Groovy-based SmartThings, OpenHAB programs) and translate them into vendor-independent trigger and action-based policies (as shown in Figure \ref{figure:dynamic_trigger_action_policy_graph}). Similarly, the {\it MUD profiles}\footnote{Currently MUD-profiles are limited to support ACL-based traffic filtering rules. It does not support dynamic trigger and action-based policies yet.} and translated into vendor-independent graph-based ACL policies (as shown in Figure \ref{figure:static_dynamic_acl_policies_graph}). This approach allows each of the functional component of \secintent i.e., composition engine to operate seamlessly on vendor-independent (VI) model. As shown in Figure \ref{figure:vise_architecture}, \secintent generates the vendor-independent (VI) graph-based policies using following key functional modules:


\begin{packeditemize}

\item
{\it ANTLR parser generator}: As a first step, we develop the lexer and parser grammar (.g4) files required to translate the vendor-specific IoT apps to an abstracted intermediate representation (i.e., Abstracted LISP Representation (ALR) and Abstracted Tree Representation (ATR) formats) with ANTLR module. 
ANTLR parser generator uses the Abstracted LISP representations (ALR) of the IoT Apps for performing the code sanity analysis (discussed in Section \ref{sec:code_sanity_analysis}). Note, both the ALR and the ATR representations of IoT apps are not exposed to end users, only the outcome of VI graph generator (\ballnumber{2}) will be exposed to users and composition engine.

\item 
{\it VI Graph generator}: In the next step, automation rules represented in the Abstracted LISP syntax representation are consumed by the VI Graph-generator module to translate it to vendor-independent graph-based policies (as shown in Figure \ref{fig_vise_translation_policy_p2}). We built (.map) file, which is used for maintain necessary mappings between different vendor-specific automation/policy attributes and vendor-independent graph attributes. These vendor-independent attributes (or labels) are used in the construction of final vendor-independent graph-based representation. The vendor-independent graph-based specifications outcome of \secintent (\ballnumber{2}) is captured using networkx python library for maintaining the policy graphs, required for composing the policy graphs to detect the conflicts (discussed in Section \ref{sec:composition_security_analysis}).  

\end{packeditemize}



\subsection{Code Sanity Analysis}\label{sec:code_sanity_analysis}
Currently, IoT administrators use both the vetted and unvetted IoT apps from the market place and customize them to cater their specific automation needs. Though, Groovy-based SmartThings market apps are heavily vetted for sanity, but still there wide range of unvetted SmartThings apps that are available online for use. Directly customizing the IoT apps without effectively analyzing them or testing could lead to {\it incoherent} automation. This problem is also highly prevalent with other programming specifications such as OpenHAB and IFTTT. 

The most prominent code sanity bugs that are reported today could include -- {\it Undefined \& Unused  variables and structures}, {\it improper If/Else closures}, {\it missing quotes}, and {\it empty If/Else blocks}, and {\it empty definition}, which \secintent verifies for. For example (as shown in Figure \ref{figure:code_sanity_undefined_unfererenced}), OpenHAB automation framework allows IoT programs to reference undefined variables and functions, and similarly not reference to the variables and functions that are defined in the IoT app. Though, in both the cases OpenHAB automation framework allowed IoT programs to run successfully with out reporting any compile time errors, but resulted in run-time violation or device misbehavior~\cite{openhab_missing_quotes,openhab_missing_if_braces}. Other code sanity bugs that leaves IoT infrastructure vulnerable are illustrated in Figure \ref{figure:code_sanity_undefined_unfererenced}. In these examples, the IoT devices are not allowed to execute action \enquote{or} did not allow the device to transition to new state, which could lead to either safety or security vulnerability in the IoT infrastructure (as highlighted in Figure \ref{figure:code_sanity_undefined_unfererenced}).

\begin{figure*}[t!]
\vspace{-20pt}
\begin{subfigure}[b]{0.24\textwidth}       
\includegraphics{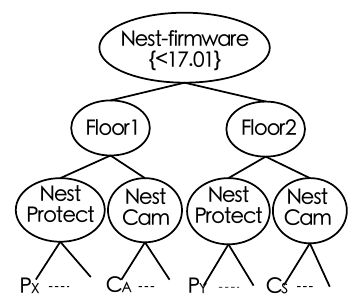}
\caption{Nest device's firmware version $<$17.01 specific to location \& type.}\label{nest_version_1701}
    \end{subfigure}
    \hfill
\begin{subfigure}[b]{0.23\textwidth}       
\includegraphics{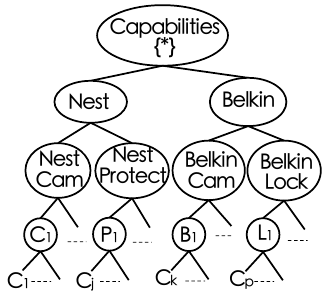}
\caption{Capability-specific abstraction for Nest \& Belkin devices.}\label{fig:nest_devices_capabilities}
\end{subfigure}
\hfill
\begin{subfigure}[b]{0.23\textwidth}       
\includegraphics{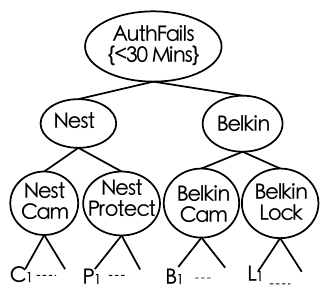}
\caption{Vendor-specific authentication fails in last 30mins.}\label{auth_fail_30mins}
\end{subfigure}
\hfill
\begin{subfigure}[b]{0.23\textwidth}       
\includegraphics{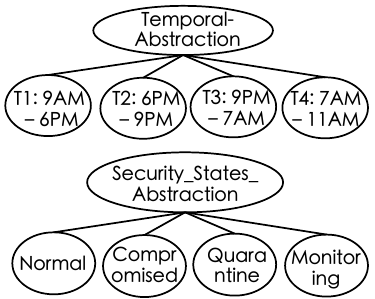}
\caption{Temporal and security state abstractions Trees.}\label{fig:temporal_security_abstractions}
\end{subfigure}
\vspace{-8pt}
\caption{\small Automatically constructed tree-based abstractions for the smart campus IoT infrastructure: (a) captures devices as leaf nodes and Abstraction, (b) captures device capabilities and property values as leaf nodes, (c) captures the dynamic security states (e.g., authentication) specific to IoT devices, (d) Temporal and Security State Abstractions of IoT infrastructure.}\label{fig:sample_network_abstractions}
\vspace{-10pt}
\end{figure*}

\secintent detects such code sanity bugs 
using the ALRs provided by the ANTLR parser generator module i.e., before giving it as input to the VI graph generator module). 
In case of the {\it defined-but-unreferenced variables and functions}, they are reported as less severe bugs as they do not break the functional capability of the IoT infrastructure. Similarly, the {\it Undefined-but-referenced} variables and functions, and empty if blocks are reported as {\it critical} bugs by the \secintent engine as these bugs could leave gap in the automation making IoT infrastructure vulnerable.

Similarly, there are other code sanity bugs, such as improper closures in which either the if and/or else conditions are not properly closed, such improper termination of code blocks leads to unexpected code flows leading to device misbehaving. For example, missing quotes ("), and missing braces (\}) as highlighted in the Figure \ref{figure:code_sanity_undefined_unfererenced} are commonly seen issues with OpenHAB and Groovy-based IoT apps, while these issues are not captured during compile time leading to run-time violations and IoT program malfunction~\cite{openhab_missing_quotes, openhab_missing_if_braces}. The code sanity and the bug severity assigned to each of the IoT apps are used in later stage to further detect the potential run-time violations present in the IoT apps (discussed in Section \ref{sec:security_analysis}).




\subsection{Infrastructure Abstractions \& Graph-based Specification}\label{sec:abstractions_graph-based-specification}

As discussed in Section \ref{sec:vi_spec}, \secintent supports mechanism to translate the vendor-specific IoT Apps into vendor-independent graph-based specification. In addition, \secintent also supports simple drag and drop-based user interface that allows administrators to directly and intuitively capture their policies in a vendor-agnostic manner using nodes of infrastructure abstraction trees, as discussed below. 




\subsubsection{Tree-based Abstractions}\label{sec:tree-based-abstractions} 
The {\it Infrastructure abstraction engine} automatically builds necessary abstraction trees required for graph-based policy specification.
The abstraction engine continuously churns data from local IoT infrastructure and cloud data sources of IoT devices using device and vendor-specific APIs for building the abstractions~\cite{nest_iot_cloud_apis,samsung_iot_cloud_apis}. The data extracted from IoT device's data sources as text and log messages are translated into data tables by the data-source-driver engine that we designed for each vendor's data format.

Generation of such abstractions allow administrators to delegate the responsibility of policy specification to sub-administrators. For example, it is now possible for a global building/campus administrator to assign Floor1 and Floor2 responsibilities to different sub-administrators. Also, \secintent allows abstraction trees to be built on physical or logical grouping of device (e.g., Figure: \ref{nest_version_1701} \& \ref{fig:nest_devices_capabilities}). The global administrator simply uses {\em abstraction mappings} to delegate infrastructure to each of the administrator. Depending on the assigned {\em abstraction mappings} the abstraction engine generates the abstraction trees. For example, to derive the abstraction tree shown in  Figure \ref{nest_version_1701} (i.e., {\it list of Nest devices of BLDG1 with firmware $<$17.01 organized as per floor and type of devices}), the abstraction-mapping required is: 

{\small
\begin{verbatim}
abstraction-tree{"Nest-Firmware{<17.01}"} = 
    location{BLDG1}.floors{*}:              
    vendor-type{Nest}.device-category{*}:   
    devices{*}                              
\end{verbatim}
}

Similarly, for capturing the list of devices and their capabilities with respect to their vendor and device-types in an abstraction tree (Figure \ref{fig:nest_devices_capabilities}), required abstraction-mapping is:

{\small
\begin{verbatim}
abstraction-tree{"Capabilities{*}"} = 
    vendor-type{*}: 
    vendor-type{*}.device-category{*}:   
    devices{*}:
    capabilities{*}
\end{verbatim}
}

\vspace{3 pt}
\noindent \textbf{Representation of Abstractions}: We represent IoT infrastructure as set of \textit{infrastructure-abstraction trees}, specific to each of the administrators. The root node of the abstraction tree (i.e., abstraction tree name) uniquely represents each of the abstraction tree present inside the IoT ecosystem. The leaf nodes represents set of individual devices. Each intermediate node represents different infrastructure abstractions such as device types, device vendors, location, temporal, application details. The infrastructure abstractions trees implicitly capture the boolean \texttt{Union} operator represented as sibling nodes in the abstraction tree. Similarly, the boolean \texttt{AND} operator corresponds to the relationship between the child and parent nodes of the abstraction tree. For representation, each level of abstraction is separated with ``:'' operator, while the constraints and conditions to be imposed on to that level is represented with a ``.'' operator. 

\begin{figure*}[t!]
\vspace{-15pt}
\begin{center}
\begin{subfigure}[b]{0.52\textwidth}
\includegraphics{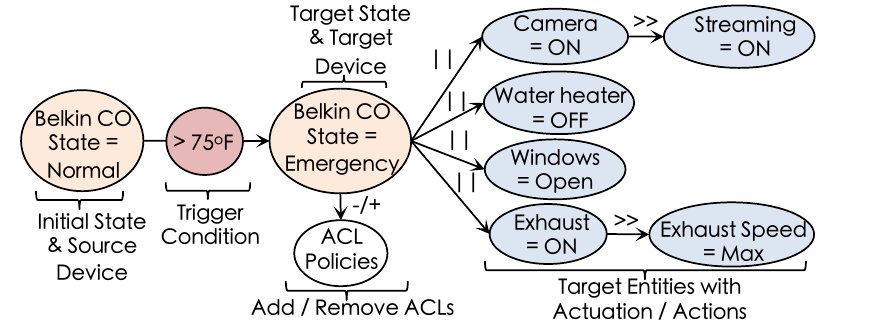}
\caption{Dynamic trigger \& action-based IoT policy specification.}\label{figure:dynamic_trigger_action_policy_graph}
\end{subfigure}
\begin{subfigure}[b]{0.46\textwidth}
\includegraphics{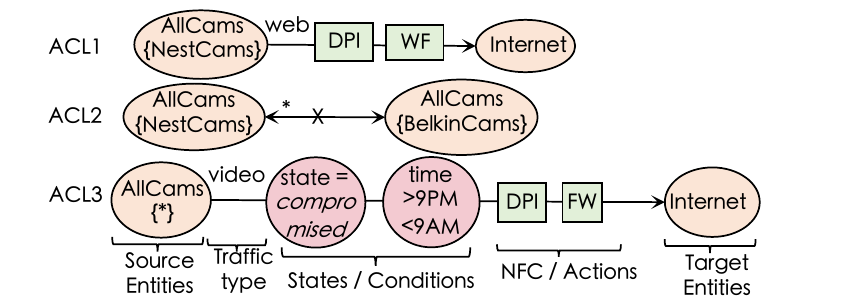}
\caption{ACL-based IoT policy specification.}\label{figure:static_dynamic_acl_policies_graph}
\end{subfigure}
\vspace{-5pt}
\caption{\small Examples of Vendor-Independent Graph-based IoT Policy Specification with policy attributes highlighted in different color.}\label{figure:both_trigger_action_acl}
\end{center}
\vspace{-15pt}
\end{figure*}

\secintent supports a diverse set of abstractions, which will allow IoT administrators to capture different types of policies based on their security status, locations, vendor-type and so on.  These include: ($i$) {\it security-state abstractions}, ($ii$) {\it location-specific abstractions}, ($iii$) {\it device- or vendor-specific abstractions}, and ($iv$) {\it application-specific abstractions}. Such tree-based abstractions allow administrators to intuitively capture their automation and policy requirements. 

\subsubsection{Graph-based Specification}\label{sec:graph-based-specification} As discussed in Section \ref{sec:vi_spec}, \secintent engine translates the vendor-specific IoT Apps into two different types of policies: ($i$) {\em trigger- and action-based policies}, and ($ii$) {\em dynamic ACL-based policies}. In addition to this capability, \secintent also supports intuitive graph-based specification framework that allows IoT administrators to directly express their policy intents through simple drag and drop-based user interface (UI), which uses policy attributes from tree-based abstractions.


\vspace{3 pt}
\noindent \textbf{Trigger- \& Action-based Policies}: The trigger \& action-based policies allows the administrator to capture their policy intents for set of IoT devices that perform some predefined action in response to an triggering event. Examples of such policies represented as graph-based policies are illustrated in Figure \ref{figure:dynamic_trigger_action_policy_graph}. These types of policies can be specified as complex graph-based policies or finite state automation (FSM), where each of the node captures abstractions, device names, conditions or actions. Our trigger- and action-based policy graphs have the following format: source IoT device with associated states, set of conditions, dynamic states, associated set of events and respective action/s on target IoT devices.

The source node represents the IoT device on which the event is received. Rest of the nodes represent conditions and state of the IoT infrastructure, except for nodes that have incoming edge with sequential ($>>$) and parallel ($||$) action operators associated with them. The sequential and parallel operators $>>$ and $||$ are used to specify the sequence of IoT-commands that need to be executed. The $+/-$ operators are used to add or remove the list of ACLs that are specific to the current state or condition.  

\vspace{3pt}
\noindent \textbf{ACL-based Policies}: This type of policies either allows or restricts the communication between devices and internet according to the dynamic infrastructure states and conditions. Examples of graph-based access control policies are shown in Figure \ref{figure:static_dynamic_acl_policies_graph}. For the trigger and action-based policy shown in Figure \ref{figure:dynamic_trigger_action_policy_graph}, we implicitly add ACLs to ALLOW traffic between Belkin CO device and other devices on which the action is enforced (i.e., Camera, Water heater, Windows, Exhaust). In reality, as the communication happens indirectly between the hub or cloud interface and the Belkin device, equivalent ACLs are added to the network. 

The starting and end nodes represents the source and destination entities of policies. The edge between the source and target node captures following properties: ($i$) set of network functions through which the traffic should traverse, i.e., network function chain (NFC), ($ii$) conditions to be enforced on the traffic depending on the state, and ($iii$) actions to be taken on the traffic between the source and destination entities. With both the ACLs and trigger-action-based policies, the states and conditions are represented as nodes along the path for simplified design in the drag-and-drop-based graph specification framework. Alternatively, these could be represented as simple edge properties. The $\,\to\,$ and {\small $-$x$\,\to\,$} arrowed-lines represents the action (i.e., ALLOW/DENY) on the traffic. The specification syntax, keywords, attributes, symbols and operators used by \secintent for processing the graph-based policies i.e., ACL-based policies and trigger and action-based policies are discussed in Section \ref{sec:specification_syntax}.

\subsubsection{Graph-based Policies Specification Syntax}\label{sec:specification_syntax} 
\begin{figure}[h!]
\begin{subfigure}[h]{0.48\textwidth}        \includegraphics{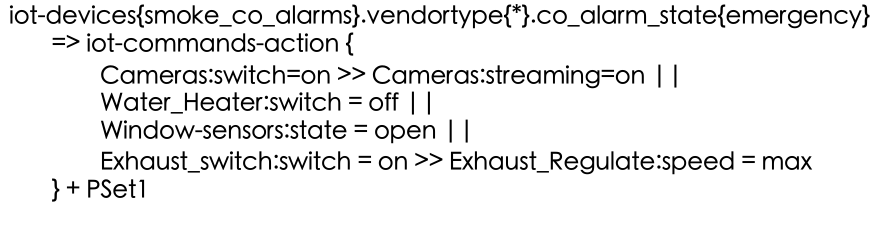}
\vspace{-21pt}
\caption{Equivalent syntax for dynamic trigger--action IoT policies.}\label{figure:dynamic_trigger_action_policy_syntax}
\end{subfigure}

\begin{subfigure}[h]{0.48\textwidth}   
\centering
\includegraphics{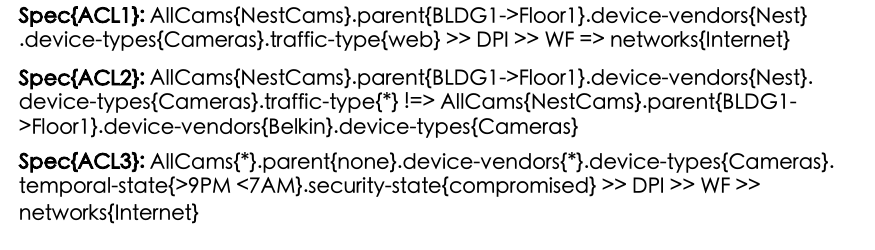}
\vspace{-20pt}
\caption{Equivalent policy specification syntax of ACL-based policies (ACL1 -- ACL3) illustrating different properties.}\label{figure:static_dynamic_acl_policies_syntax}
\end{subfigure}

\caption{Examples of ACL-based \& Trigger and Action-based IoT policy specification syntax.}\label{static_dynamic_acls_specification_both}
\end{figure}

An equivalent specification syntax is required for capturing each of these graph-based policies in the backend. For example, in the ACL-based policy specification syntax, the permissions to communicate between source and target nodes is specified using action attributes $=>$ (i.e., ALLOW) and $!=>$ (i.e., DENY) symbols. The sequence of network functions through which the traffic from a specific source node should traverse to reach the target node or the sequence of actions to be taken is specified using the sequential or parallel operators (e.g., Firewall $>>$ DPI, FW $||$ LB, Exhaust$=$ON $>>$ ExhaustSpeed$=$High). Note that the sequence and parallel operators are used for traversal of a set of middleboxes in the case of ACL-based policies. In case of trigger-action-based policies, it is used to represent the sequence of actions. Also, in the trigger-action-based policy specification syntax, actions are captured using the {\tt \small iot-commands-action} keyword. 

\begin{table}[h!]
\vspace{-5pt}
\footnotesize
\begin{tabular}{p{0.56in}|p{1.05in}|p{1.37in}} \hline
\textbf{Type} & \textbf{Symbol} & \textbf{Definition}  \\ \hline
Policy specification keywords & location\{\}, devices\{\}, device-type\{\}, device-vendors\{\}, parent\{\}, traffic-type\{\}, source-node\{\}, target-node\{\}, source-state\{\}, target-state\{\}, etc., & Keywords for capturing policy attributes and the properties of the IoT infrastructure. \\ \hline

Sequence \& Precedence Operators & $>>$ (serial or precedence), $||$ (parallel), $\,\to\,$ (flow / action sequence) & Operators to specify the sequence of operations to be carried out in policy. \\ \hline

Conditional Operators & $!$, $=$, $<$, $>$ & Operations used to specify dynamic conditions in the policy. \\ \hline

Composition Operators & policy-add{} ($+$), policy-remove{} ($-$) & Operators for adding and removing policies from the existing list of policies.\\ \hline

Action attributes / Keywords & iot-commands-action, $=>$ (ALLOW),  $!=>$ (DENY) & Attributes or keywords used to specify the actions to be taken in the policies. \\ \hline

\end{tabular}
\caption{\small List of keywords, attributes, operators and symbols used in \secintent policy specification syntax.}\label{table:policy-attributes}
\vspace{-15pt}
\end{table}

\secintent allows duplicate node names to be captured as part of same or different policy abstraction trees, which could lead to discrepancies when policies are specified using these nodes. To overcome this problem, we maintain the root nodes names of policy abstraction trees be unique across the IoT infrastructure. We use the \enquote{\textit{parent}} keyword to uniquely identical nodes across different abstraction trees. The $\,\to\,$ operator is used with {\it parent} to define the path of the source entity from its \textit{parent}. For example, following reserved keywords are used to group set of devices \textit{device-types}\{\enquote{{\tt \small Cameras}}\}, \textit{vendor-types}\{\enquote{{\tt \small Nest}}\}, \textit{location}\{\enquote{{\tt \small BLDG1$\,\to\,$Floor1}}\}, and 
\textit{smoke\_co\_alarm\_state} \{\enquote{{\tt \small emergency}}\}. Other keywords, operators, and symbols used in \secintent are listed in Table \ref{table:policy-attributes}. For the trigger and action-based illustrated in Figure \ref{figure:dynamic_trigger_action_policy_graph} the equivalent syntax is shown in Figure \ref{figure:dynamic_trigger_action_policy_syntax}. Similarly, for the ACL-based policies illustrated in Figure~\ref{figure:static_dynamic_acl_policies_graph}, the equivalent policy specification syntax is shown in the Figure \ref{figure:static_dynamic_acl_policies_syntax}.



%
%
\section{Graph Composition \& Security Analysis}\label{sec:composition_security_analysis}
Manually detecting and effectively resolving the compile-time conflicts and run-time violations among automation policies even in small-scale IoT deployments is an arduous and error-prone process. Hence, \blue{\secintent effectively addresses this challenge in this section.} 

\subsection{Rogue policies}\label{sec:composition_rogue_static_conflicts} 
\secintentnospace's \textit{abstraction trees} and {\it graph-based specification} allows the administrators to specify the policies explicitly using the abstraction trees exposed to each of the IoT users or administrators. This approach of isolating and assigning explicit infrastructure abstractions to each admin, allows \secintent to prevent admins from specifying policies on the infrastructure they do not own. However, it should be noted that the policies that are directly specified using policy abstraction syntax (as discussed in Section \ref{sec:specification_syntax}) i.e., bypassing the web-based user interface provided by graph-based policy specification is also detected for rogue policies by the policy composition engine and updated to the administrators as violations. 

For detecting {\it rogue policies} in syntax-based policy specification, the policy composition engine extracts the source and target nodes from the specified policy specification syntax and verifies if both the nodes belongs to the \textit{policy abstraction trees} owned by that administrator. If the policy attributes used in the policy specification are part of the abstraction trees assigned to that admin, then the policies are allowed to perform actual policy composition to further detect other conflicts and violations (discussed in Section \ref{sec:security_analysis}). The policies violating this step is reported as {\it rogue policies}. 

\subsection{Graph-based Composition 
}\label{sec:graph-based-composition}
The policy composition in \secintent seeks to provide the following capabilities: ($i$) proactive composition of policies for detecting conflicts in compile time rather than detecting at run time, ($ii$) automatic resolution of conflicts using {\em precedence} mechanism, and ($iii$) {\em incremental composition} that allows trigger-action policies to adapt to the environment at runtime. However, allowing run-time policy composition imposes stringent latency constraints. Hence, policy composition time has to be minimized.

\texttt{\secintentnospace}'s policy composition procedure involves following key steps. First, the {\em normalization} step brings all the policies specified by various administrators, using different abstraction trees, to a common abstraction level for identification of contradictory and duplicate policies.  The second step, finds contradictions among normalized policies by running composition engine and resolves the conflicts using precedence rules. Unresolved conflicts are flagged to the administrator.


\subsubsection{Proactive \& Incremental Composition}\label{sec:incremental_composition}
The goal of policy composition mechanism is to produce {\it conflict-free composed policy graph} (example shown in Figure \ref{fig:final_composed_policy}) that is derived by composing together overall vendor-independent graph-based policies (i.e., either generated from IoT apps or specified directly by policy administrators). In the final composed graph, the source and target entities represent the devices or group of devices onto which policies are executed. The edges capture the set of conditions for policy activation and the corresponding actions. 

As a first step of policy composition, \texttt{\secintentnospace}'s normalization mechanism identifies the common abstraction level to which all policies specified by administrators need to reduced for conflict detection. If policies are specified using abstraction nodes at different levels of an abstraction tree, composing policies without normalization is an infeasible task. For example, consider two policies specified using the abstractions BLDG1 and Floor2 (homogeneous abstractions) or using BLDG1 and NestCams (heterogeneous abstractions). Here, we do not know if the  Floor1 node and Nest node have any relation (i.e., subset, superset, or overlapping hosts or devices) for composing them together into a single policy graph. We resolve this issue by ($a$) automatically deriving relations across non-leaf nodes of different abstraction trees and maintain these relations as mappings i.e., capturing the details of set of devices or hosts that belongs to each of these nodes of abstraction trees, and ($b$) choosing an optimum level to which the nodes used in the policy specification need to be normalized. 

\vspace{3pt}
\noindent \textbf{Normalization}: A naive normalization approach is to bring all the policies to the bottom most level, i.e. leaf node level which captures the device-specific details for performing composition. This approach increases complexity along two dimensions: ($i$) It brings down all  policies to the bottom most level even though very small number of policies might need it, exponentially increasing the composition times, and ($ii$) normalizing all the source and destination policy nodes to bottom most level (i.e., to the level of leaf nodes) also increases the enforcement complexity, due to increase in the number of rules required to enforce the policies. Therefore, \secintent finds an intermediate level i.e., Enforcement Level (\texttt{ELevel}) for each policy abstraction tree for effectively normalizing the policies. The enforcement level (\texttt{ELevel}) is chosen in such a way it would allow the policy enforcement engine to directly compile/translate these composed policies to enforceable rules. 

Mathematically, we represent the choice of enforcement level in the following way. Let $K_i$ be the abstraction level using which a policy $P_i (i = 1,\ldots,N)$ is specified. Then, policy $P_i$ can be normalized any abstraction level $A_i \geq K_i$.
Thus, we aim to normalize to a level $A$ such that $A \geq A_i \, \forall i=1,\ldots,N$.
However, the number of nodes increases with an increase in $A_i$. Thus, our objective is to choose the minimum value of $A$ that satisfies the above set of constraints, i.e.
\begin{equation}
\text{Minimize } A \text{ subject to: }A \geq K_i \ \forall i=1,\ldots,N.
\end{equation}



Algorithm \ref{algo:graph_composition} describes the graph-composition algorithm, which accepts: ($i$) a list of graph-based policies that are normalized, and ($ii$) an empty graph as input, which is used for storing the composed policy graph. Each of the input policy is composed with the graph for conflict identification and resolution. The list of input policies are iteratively checked for an identical or conflicting policy. First each policy's source node is compared with the source nodes of the composed policy graph.  Next, the target node of the input policy is compared with the target nodes of the composed policy graph. Finally, the conditions and action attributes present along the edge of the input policy are compared with the edge attributes of the composed policy graph. In this process, the composed policy graph is updated with non-conflicting policy attributes. 

The algorithm attempts to resolve conflicts by checking whether a precedence rules exists for any of the policies. Duplicate policies are tracked separately and discarded from composition. If it finds a conflicting policy that cannot be resolved with precedence, the policy is dropped from the graph (Lines 11-13) and notified to administrator. If it finds that the source node and the destination node both exist and no conflicts are possible, the policy is then added to the graph (Lines 14-16). Otherwise, it creates new nodes and then adds the policies as edges to the graph (Lines 17-23).

\begin{algorithm}[t!]
\footnotesize
\caption{Graph-based Policy Composition}\label{graph_composition}
\label{algo:graph_composition}
\begin{algorithmic}[1]

\State $L \gets \text{list of normalized policies to be added}$
\State $s(p) \gets \text{source node of policy } p$
\State $t(p) \gets \text{destination node of policy } p$
\State $a(p) \gets \text{action of policy } p$
\State $b((s, t)) \gets \text{action of edge } (s, t)$
\State $G \gets \text{Composed graph}$
\ForAll {Policy $p \in L$}:
	\If {$s(p) \in G \ \& \ t(p) \in G$}
		\If {$(b(s, t) \in G \ \& \ a(p) == b(s,t)$}
				\State {Discard $p$} \Comment{Duplicate Policy}
		\ElsIf {$b(s, t) \in G \ \& \ a(p) \neq b(s, t)$ }
        	\State {Apply $b(s, t)$ or $a(p)$ based on precedence}
			\State {Raise conflict alert if policies have equal precedence}
		\ElsIf { $b(s,t) \notin G$}
			\State {Create $b(s,t)$ from $p$}
			\State {Add $b(s,t)$ to $G$} \Comment{Add policy}
		\EndIf
	\Else
		 \If {$s(p) \notin G$}
			 \State {Create $s(p)$}
		 \EndIf
		 \If {$t(p) \notin G$}
			 \State {Create $t(p)$}
		 \EndIf
		\State {Create $b(s, t)$ from $p$}
		\State {Add $b(s, t)$ to $G$} \Comment{Add policy}
	\EndIf
\EndFor
\Return $G$
\end{algorithmic}

\end{algorithm}

To analyze the algorithmic time complexity, we need to evaluate the cost of iterating over the complete list of policies ($L$), and adding them to the composed Graph $G$. With the addition of each policy to the composed policy graph $G$, the composition engine checks for the existence of conflicts. The major time complexity of the algorithm lies with the iteration of the policies O($L$) over the list of all source nodes $L_s$ in the composed graph $G$, and then comparing the policy's source node $s(p)$ to composed graph's source nodes $S(G)$. $L_e$ is the list of edges for which the edge-properties overlap with the policy's edge among the overlapping source nodes. Also, $L_t$ is the list of target nodes of the edges of $s(p)$ that actually overlaps with the target node of the policy $p_i$. Therefore, the overall worst-case complexity is: $O(L*L_s*L_e*L_t)$. 

\blue{For reducing the composition complexity we employ hashing mechanism and caching technique: ($i$) the $m$ host entries of $s(p)$ are hashed as key-value pairs and the host entities of $S(G)$ are looked up in the hash for the existence of the $n$ hosts, reducing baseline complexity will be reduced to: $O(L * L_s * (m+n))$. ($ii$) caching the comparison calculation outcome as key-value pairs ($s(p)$:$S(G)$) in hash table reduces the overall baseline complexity to $O(L*L_s)$.}  


\subsubsection{Precedence}\label{sec:precedence}
Precedence rules are used to resolve conflicts among competing policies specified at different levels. {\it Administrator-level precedence} evaluation is based on the scope of authority of the policy author. For example, a campus-level administrator in a smart campus may be granted precedence over a building administrator; 
{\it Action-level precedence} allows for explicit prioritization in action invocation. For example, for IoT traffic's ACL-based policies, Drop $>$ Allow $>$ Quarantine $>$ Redirect can be used as the precedence hierarchy. Similarly, in the case of trigger-action-based policies, when the smoke detector is in {\tt \small fire-alarm} state, the action turn OFF heating is given higher precedence than turn ON heating. {\it Custom precedence} enables policy attributes (e.g., {\tt \small user}, {\tt \small device type}, {\tt \small device state}) to be associated with precedence. 

When two policies (P1 and P2) conflict, the nodes and edges of the policies are decomposed into the {\it set of subset nodes that requires the least number of edges to represent conflict-free policies} (for optimizing the number of rules required for enforcement). Based on precedence, the overlapping nodes that result in conflict are removed and the edge specific to the policy with highest precedence is retained.

\subsubsection{Incremental Updates} The dynamic characteristics of the consumer IoT infrastructure demand that the policy framework be agile in enforcing new set of rules to IoT devices. Since complete policy composition consumes time, up to a few minutes, efficient re-composition techniques are required for rapid policy response.  Hence, we use \textit{incremental policy composition} to ensure an expedient response to dynamic conflicts that arise in the network. The composition engine recomposes only the updated set of policies with the whole set of composed policies. Updating a policy from the composition graph involves first deleting the policy  from the graph, and then inserting a modified version. Deleting a policy requires one to remove the edges that belong to the policy from graph.  However, the composition procedure might have removed portions of other policies that had a higher precedence during conflict resolution. Hence, these lost portions must be returned. 

\begin{figure}[t!]
\centering	
\includegraphics{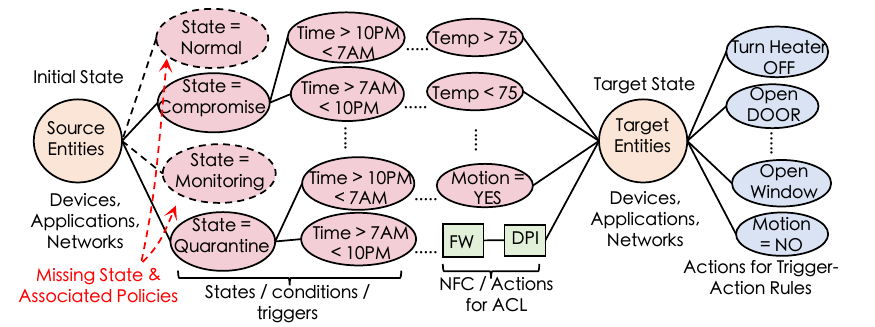}
\vspace{-20pt}
\caption{\small Sample outcome of graph-based policies composed together representing final composed policy graph. Possible missing states are highlighted in the figure with dotted lines.}\label{fig:final_composed_policy}
\end{figure}

Incremental policy composition reduces the time to react to dynamic state changes and enforce new rules under the following scenarios: ($i$) {\em Scenario 1}: New polices are added or removed from the IoT infrastructure; ($ii$) {\em Scenario 2}: IoT infrastructure changes (e.g., device location updates, new devices added or existing devices removed); and ($iii$) {\em Scenario 3}: IoT device-state changes (e.g., from {\em normal} to {\em compromised}). Also, for run-time composition \secintent enumerates all the possibilities to pre-calculate the possible outcomes. Hence, for run-time enforcement \secintent simply identifies the outcome specific to the event and simply executes it, which can be carried out in sub-second latency, much faster than incremental composition.



\subsection{Security Analysis \& Policy Enforcement}\label{sec:security_analysis}
In this section, we describe about how we use the graph-based composed policy graph to perform security analysis for detecting the bugs and violations. We discuss about the conflict-free policy enforcement by reconciling policies to device-specific rules and automatic policy inference.

\subsubsection{Gap Analysis}\label{sec:gap_analysis}
Gap in the automation could leave the IoT infrastructure in a state that is either unstable or unpredictable in its behavior. Such dangling state could make IoT infrastructures vulnerable to attacks. For example the policies $S_{5}$, $S_{12}$ and $S_{14}$ shows the gap in the automation with respect to its temporal and temperature conditions (as described in Table \ref{table:smart-building-polcy-conflicts}). It is evident from these policies that during 8PM --9PM (conflict) 9PM -- 9AM (Gap) the thermostat's temperature settings could not be effectively predicted. As listed in Table \ref{table:smart-building-polcy-conflicts}, similar automation gap could be visible in other types of policies specific to its spatial, security states and other environmental conditions. Therefore, identifying the gap in automation is key step towards detecting the {\it potential} bugs that might arise in the IoT infrastructure during run-time.

The \vice module traverses through the final composed policy graph to identify the missing states or conditions for which the policies are not captured (as shown in Figure \ref{fig:final_composed_policy}). 
This is achieved by enumerating and verifying if the policies exist for all the possible temporal, spatial and security conditions (e.g., states captured as part of abstraction tree's leaf nodes such as shown in Figure \ref{fig:temporal_security_abstractions}), which helps in identifying potentially missing policies i.e., gap in automation. For example, the possible security states of IoT infrastructure could be {\it Normal}, {\it compromised}, {\it monitoring} and {\it quarantine} (Figure \ref{fig:temporal_security_abstractions}). By identifying the missing states and their associated policies using the abstractions tree as reference, we can effectively detect the gap in automation. 

The completeness of the gap analysis depends on the abstraction engine's ability to extract all the possible states, conditions and infrastructure details captured as part of the abstraction tree. For example, the temporal, spatial and security abstractions that are auto-generated by the abstraction engine could be verified by user for its correctness, allowing the administrator to add the missing states as leaf nodes to the abstraction trees. 


\subsubsection{Loops in Automation}\label{sec:loop_chains}
In general detecting chains and loops within the automation rules is essential to detect the potential conflicts and violations that might arise during run-time, which are rather challenging to be detected at policy compile time. From the composed policy graph, we detect the loops as follows: ($i$) We check for existence of paths with in a composed policy graph that has more than one trigger and action pair (i.e., chaining among policies). ($ii$) Check if any of the actions along the path triggers back any of the events with in the chain. ($iii$) Check if any of the actions along the path triggers any of the events with in the chain resulting in taking different action, which results in ambiguity and continuous toggling of states and actions. Loops among the policies with continuously toggling actions are marked as conflicts, while the identified policy chains are marked for potential violations (e.g., policies $S_{5}$, $S_{6}$, $S_{7}$ and $S_{10}$ in Table \ref{table:smart-building-polcy-conflicts}). 
These policies when realized together will result in either unintended behavior, continuous toggling of actions or might result in unsafe state i.e., leaves door locked or opened during unanticipated instance of time resulting safety violations or sets unintended temperature conditions in home. 

\subsubsection{Potential Conflicts}\label{sec:potential_conflicts}
Existing tools lacks support for proactively detecting the policy violations that might arise during run-time. Consider for example policies $S_{10}$ and $S_{15}$, which results in run-time violations. These two policies does not have any thing in common except the actions taken by them i.e., $S_{10}$ closes windows in case of specific outdoor temperature, while $S_{15}$ opens the windows in case of raining and humidity. 
Therefore, identifying the run-time conflicts is a challenging task with such policies. As a straw-man solution one could mark all the policies that has conflicting actions and lack of specific temporal attributes among these policies as potential violation. Such approach will result in generating vast false positive. 
Exposing large number of false positives to novice smart home users could prevent them from using the automation framework, rendering these tools useless.


Therefore, it is essential to identify a means by which the potential violations are ``proactively'' detected and fine-tuned to avoid run-time violations. To identify {\it potential} run-time violations
\secintentnospace, checks for {\it mutually exclusiveness}\footnote{Two policies are mutually exclusive, when no two events, conditions or states among these policies are not related to each other or can co-exist i.e., occur at the same instance of time.} among the policies. For detecting {\it potential run-time violation}, we build relation among all the events, states and environmental conditions that are part of the IoT infrastructure and cluster the events. By clustering the events, states and conditions, we will be able to effectively detect if parameters among any two policies are mutually exclusive or not. To achieve this, we encode our policies~\cite{multi_label_binarizer} (i.e., captured as policy tuple: <{\it source-node}, {\it source-state}, {\it edge states/conditions}, {\it target-node}, {\it target-state}, {\it actions}>) and use k-means clustering with elbow method~\cite{k-means_clustering_states} for effectively clustering different types of triggering conditions and associated states. If the policy states and conditions are not mutually exclusive (i.e., on the basis of cluster distance) and pose conflicting actions are considered to generate potential violations during run-time. 

In addition, for the list of policies that are marked to generate potential run-time violations, we develop the severity metric to prioritize the reported violations, which  allow administrators to fix the violations in the order of their severity. We take into consideration the lack of temporal attributes, code sanity and relations derived among states as metric (i.e., aggregated sum of weights derived from all these three factors) and uses z-score and correlation ranking mechanism to evaluate the severity ranking of any potential violation~\cite{z-ranking_statistical_engler, correlation_ranking_engler}. 




\subsubsection{Policy Reconciliation and Inference}\label{sec:policy_enforcement}
As discussed in Section \ref{sec:vi_spec}, the policy mappings (i.e., mapping between the IoT app and the respective graph-based policy) 
maintained by the \secintent engine helps in effectively enforcing the policies after the conflicts are detected and resolved. The conflict-free composed policy graph along with these mappings are provided as input to the policy reconciliation engine (Figure \ref{figure:viscr_system_arch}). We develop APIs that effectively leverage this association and reconcile composed policy graph to device-specific IoT apps for enforcement. The policies that are directly provided as graph-based policy input to the \secintent are reconciled to device-specific rules and distributed on to network of IoT devices on the basis of: ($i$) vendor-type, ($ii$) device location, ($iii$) source devices that are generating the triggering events, and ($iv$) destination nodes on which the actions need to be enforced.

In case of multiple devices that can accommodate same rule, the rules are distributed uniformly preventing any of the IoT devices from getting overloaded~\cite{jen_rule_placement,adoptable_rule_placement}. Upon choosing the right set of devices on which the rules are required to be placed, the reconciliation engine translates it device-specific rules for configuring or programming the specific device. 

\begin{figure}[t!]
\centering	
\includegraphics[width=1.0\linewidth]{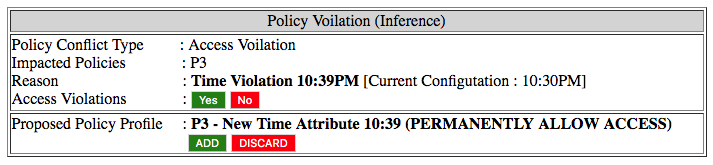}
\vspace{-20pt}
\caption{\small Inferred policy proposed to administrator for fine-tuning existing policy (P3). Yes to accept it as true positive and No to reject false positive. Propose new policy for fine-tuning the policy P3 i.e., propose new time to access the main door.}\label{fig:policy_inference}
\end{figure}

As a strawman solution, we infer new policies in accordance with dynamic changes in the behavior of IoT ecosystem for fine-tuning the automation. For example, a newly inferred policy that is proposed to an IoT administrator for fine-tuning the existing main door access policy (P3) is shown in Figure \ref{fig:policy_inference}. We plan to enhance the technique to efficiently infer new policies from the IoT ecosystem and propose right set of policies to IoT administrator for fine-tuning the existing policies `or' resolving the run-time violations (i.e., with reduced or no false positives) as part of our future work (\S \ref{sec:discussion_limitations}).

\section{Prototype \& Evaluation}\label{sec:prototype_evaluations}

We developed the complete prototype implementation of \secintent in Python and integrated it with our \secintent policy specification dashboard. 
We implemented following components in \secintentnospace:

The abstraction engine of \secintent is integrated with vendor-specific cloud data sources for extracting dynamic states and configurations of IoT devices. 
We developed data-source drivers for IoT devices 
that translates logs and text data into data tables. The data source drivers of \secintent is integrated with the Congress engine~\cite{congress_arch}, which is enhanced to generate datalog rules required for automatically generating tree-based infrastructure abstractions based on the {\em abstraction mappings} supplied by administrators. The data-push option provided by each of the vendor's cloud data-sources allows \secintent to capture the log messages specific to dynamic updates in the IoT network. 

The graph-based policies specified using the abstraction tree nodes are translated first into equivalent vendor-independent specification syntax and ultimately into a graph dictionary. We used the \texttt{networkx} Python library\cite{networkx_nxgraph} for capturing and building policy graphs. For visualizing graph-based policies and the composed policy graph, we built the \secintent policy specification dashboard integrated with networkx~\cite{networkx_nxgraph} and GraphViz library~\cite{graphviz}. 
\secintent maintains necessary mapping between the IoT apps and graph-based policies, which helps the reconciliation engine to effectively translate composed graph-based policies back to IoT programs during policy enforcement (as discussed in Section \ref{sec:policy_enforcement}). 

The composition engine captures the policies in \texttt{networkx}-based graphs, runs normalization and composition algorithms to detect the conflicts and resolves them using precedence rules that are maintained as key value pairs. These conflict-free policies are further analyzed for other bugs and violations. Finally, the composed conflict-free policies are translated to enforceable rules i.e., IoT apps and device-specific configurations. The \secintentnospace's policy composition outcome is interfaced with the \secintent specification dashboard for fine-tuning the policies before enforcement.

\subsection{Evaluation}

\begin{figure}[t!]
\includegraphics{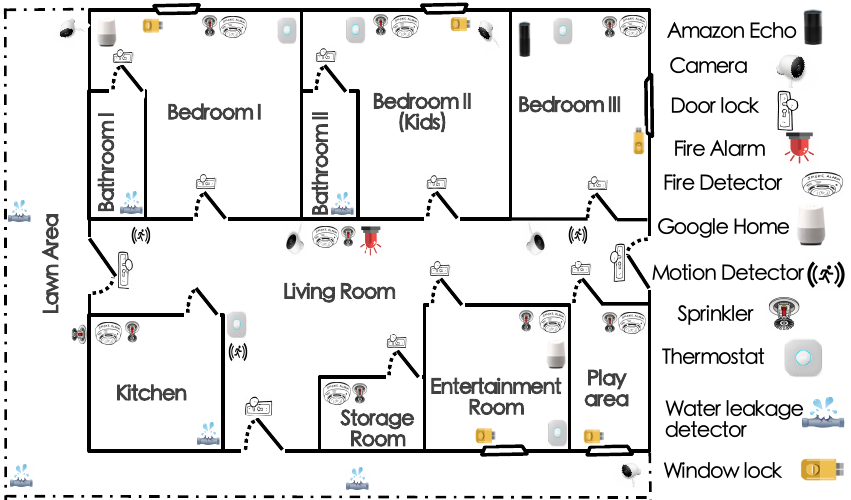}
\caption{\small Simulated Smart building IoT Infrastructure for \numofapps IoT Apps using Groovy-based SmartThings, OpenHAB, IFTTT-based Applets and MUD profiles with 30 different devices from 8 different vendors. For simplicity only 11 different types of devices are shown here.}\label{figure:simulated_testbed}
\end{figure}

\noindent{\bf Testbed.} 
We used simulated smart building IoT infrastructure with \numofapps IoT apps or automation policies framed with 30 different types of consumer IoT devices from 8 different vendors framed for multiple floors of the building. For brevity few policies (Table \ref{table:smart-building-polciies}) and simulated smart building view of single floor (shown in the Figure \ref{figure:simulated_testbed}). The \secintent module is run on a Dell R710 server with 48GB RAM, 12 cores (2.6GHz) with Ubuntu 4.4.0-97-generic kernel.

\vspace{3pt}
\noindent{\bf Datasets.} We use IoT market apps (i.e., both vetted and unvetted), extracted from vendor marketplaces, and publicly available data sources for the smart-home and smart-campus use cases~\cite{iot_test_bench,smart-home-data,smart_homedata1,sfo_dataset_all,smart-city-survey}, also consolidated in our \secintent repository~\cite{viscr_secintent_dataset}. 
We have built following three datasets for our experiments:

%
%
\begin{table}[t!]
  \footnotesize
  \centering
\begin{tabular}{|p{0.29in}|p{2.8in}|} \hline
\Centering\bfseries Policies & \Centering\bfseries Smart Building Policy Description \\
\hline
$S_{1}$ & Any time fire is detected, turn on sprinklers and cameras, and shut down all locks (doors and windows) \\ \hline
$S_{2}$ & Any time water leak is detected on office walls, cut the water supply \\ \hline
$S_{3}$ & From 10PM to 7AM keep the outer doors and windows locked \\ \hline
$S_{4}$ & From \blue{6PM to 11PM} keep the bedroom1 window \blue{unlocked or open} \\ \hline
$S_{5}$  & From 8PM to 9PM set the thermostat to 65$^\circ$F in bedrooms (kid) \\ \hline
$S_{6}$ & Between 6PM to 10PM (i.e., till \blue{11PM}) keep the main doors unlocked \\ \hline
$S_{7}$ & If main doors and windows are open for more than 5 minutes, turn OFF the heating/cooling in that room to prevent energy wastage \\ \hline 
$S_{8}$ & If motion is detected then turn the camera ON \\ \hline
$S_{9}$ & If outside humidity is $<$40\% and $>$50\%, make sure outer doors and windows are locked while maintaining humidity between 40 -- 50\% inside. \\ \hline
$S_{10}$ & When outside temperature is between 60-75$^\circ$F open the windows and turn OFF cooling \\ \hline
$S_{11}$ & From 10PM to 7AM if any of the outer doors and windows are unlocked, trigger the alarm and SMS notify \\ \hline
$S_{12}$ & From 9AM to 9PM set the thermostat to 74$^\circ$F (parent) \\ \hline
$S_{13}$ & Turn bedroom II Camera OFF (or no access) after 10PM (kid) \\ \hline
$S_{14}$ & If building temperature rises above 95$^\circ$F, lock all windows and reset the thermostat to 65$^\circ$F  \\ \hline
$S_{15}$ & In case of rain and humidity $<$40\% and $>$50\% close the windows  \\ \hline
$S_{16}$ & Keep Camera ON in all rooms and access to it at any time (parent) \\ \hline

	\end{tabular}
\caption{\small Example list of Smart building policies ($S_{1}$-$S_{16}$).}\label{table:smart-building-polciies}
\vspace{-15pt}
\end{table}

%
%
\begin{table}[h!]
\footnotesize
  \centering
\begin{tabular}{|p{0.35in}|p{2.4in}|p{0.31in}|}
\hline
\Centering\bfseries Policy Conflict & \Centering\bfseries Conflict Description & \Centering\bfseries Conflict Type \\
  \hline
$S_{1}$, $S_{2}$ & An incident where fire is detected by smoke alarm ($S_{1}$), sprinkler is triggered this can be flagged as a water leakage ($S_{2}$), which results is cutting down water supply interrupting functionality of sprinkler. Thus, resulting in a fire spread. & Static \\ \hline

$S_{1}$, $S_{3}$ & When fire is detected between 10PM and 7AM by smoke alarm ($S_{1}$), all doors and windows are unlocked (open). With $S_{3}$ both exterior door and all windows must be locked (close) during this time. This can result in exterior doors and windows toggling from locked or unlocked resulting in unintended behavior. & Static, Loops \\ \hline

$S_{1}$, $S_{14}$ & If temperature raises above 90$^\circ$F, it will enforce that all windows must be locked and thermostat be set to 65$^\circ$F ($S_{14}$). Conflicts with temperature raised due to fire event ($S_{1}$). & Static \\ \hline

$S_{9}$, $S_{10}$ & With overlapping state between $S_{9}$ and $S_{10}$, i.e, it's humid and also temperature is between 60-82$^\circ$F, in such situation windows can toggle between locked and unlocked. & Static \\ \hline



$S_{5}$, $S_{6}$, $S_{7}$ & Between 7PM and 9PM the outer door and windows are locked, thus trigger for both $S_{5}$ and $S_{7}$ are valid as a result system will toggle between turning off thermostat and setting it to 65$^\circ$F. Similarly, $S_{6}$ can further intervene due to time overlap and can result in chain and again forming  a loop. & Chain, Loop \\ \hline

$S_{7}$, $S_{10}$ & When temperature outside is 60 -- 75$^\circ$F, $S_{10}$ opens the windows, which can possibly trigger $S_{7}$ given exterior doors are unlocked too. This policy chaining might result in unintended temperature conditions inside home. & Chain \\ \hline

$S_{3}$, $S_{4}$ & Rogue behaviour as $S_{3}$ is set by parent and $S_{4}$ is set by a kid for a bedroom1 window's opening. & Rogue \\ \hline

$S_{5}$, $S_{12}$ & Rogue behaviour as $S_{12}$ is set by admin1 (or parent) and $S_{13}$ is set by admin2 (or kid) on bedroom1 in setting thermostat. & Rogue \\ \hline

$S_{13}$, $S_{16}$ & Rogue behaviour as $S_{13}$ is set by parent and $S_{16}$ set by kid in accessing kid's room Camera after 10PM. & Rogue \\ \hline

$S_{5}$, $S_{12}$ & Gap in automation between 9PM -- 9AM, where thermostat condition is not specified and conflict during 8PM -- 9PM. & Gap, Static \\ \hline

$S_{14}$ & Gap as condition is not specified for temperature less than 95$^\circ$F (i.e., between 74$^\circ$F to 95$^\circ$F). & Gap \\ \hline

$S_{10}$, $S_{15}$ & When it is both raining and temperature between 60 -- 75$^\circ$F, conflicting actions arise i.e., undecidable if windows should be opened or closed & Potential Violation \\ \hline

	\end{tabular}
\caption{\small Smart building conflicts, bugs and violations detected by \secintentnospace. For simplicity few conflicts are reported for the policies described in Table \ref{table:smart-building-polciies}.}\label{table:smart-building-polcy-conflicts}
\end{table}

\noindent \texttt{\textbf{DS-1}}:  We simulate a smart building IoT infrastructure (as shown in Figure \ref{figure:simulated_testbed}) with multiple floors and automate it with \numofapps IoT apps. We use \numofapps Groovy-based IoT apps (i.e., homogeneous specifications) for evaluating static analysis-based technique (such as \soteria~\cite{soteria}) and compare with \secintentnospace. The same set of automation rules programmed with Groovy, OpenHAB, IFTTT, and MUD profiles (i.e., heterogeneous specifications) are used for evaluating \secintentnospace. With dataset {\tt DS-1}, we evaluate our policy composition engine for detecting static violations and compare it with static violation detection mechanism (e.g., \soterianospace). In addition to detecting static policy conflicts, \secintent also detects other type of conflicts, violations and bugs from the dataset {\tt DS-1} (as highlighted in Table \ref{table:smart-building-polcy-conflicts}). 

\noindent \texttt{\textbf{DS-2}}: We use the smart-home, smart-campus and smart-city abstractions dataset~\cite{sfo_dataset1,sfo_dataset2,sfo_dataset3,sfo_dataset_all,smart-home-data} for evaluating the performance of our tree-based abstraction engine in constructing the abstraction trees. 

\noindent \texttt{\textbf{DS-3}}: We have built tool to generate $\approx$20K synthetic policies emulating the dataset {\tt DS-1}, using the policy abstraction trees generated from \texttt{DS-2}. We use random sampling technique to select the source nodes, destination nodes and edge properties (e.g., dynamic environmental states, conditions, and traffic type) from the policy-abstraction trees. 
\subsubsection{Policy Abstraction} 
\begin{figure*}[tbp!]
\vspace{-20pt}
\begin{center}
\begin{subfigure}[t]{0.23\textwidth}
\centering
\includegraphics[width=1.0\linewidth]{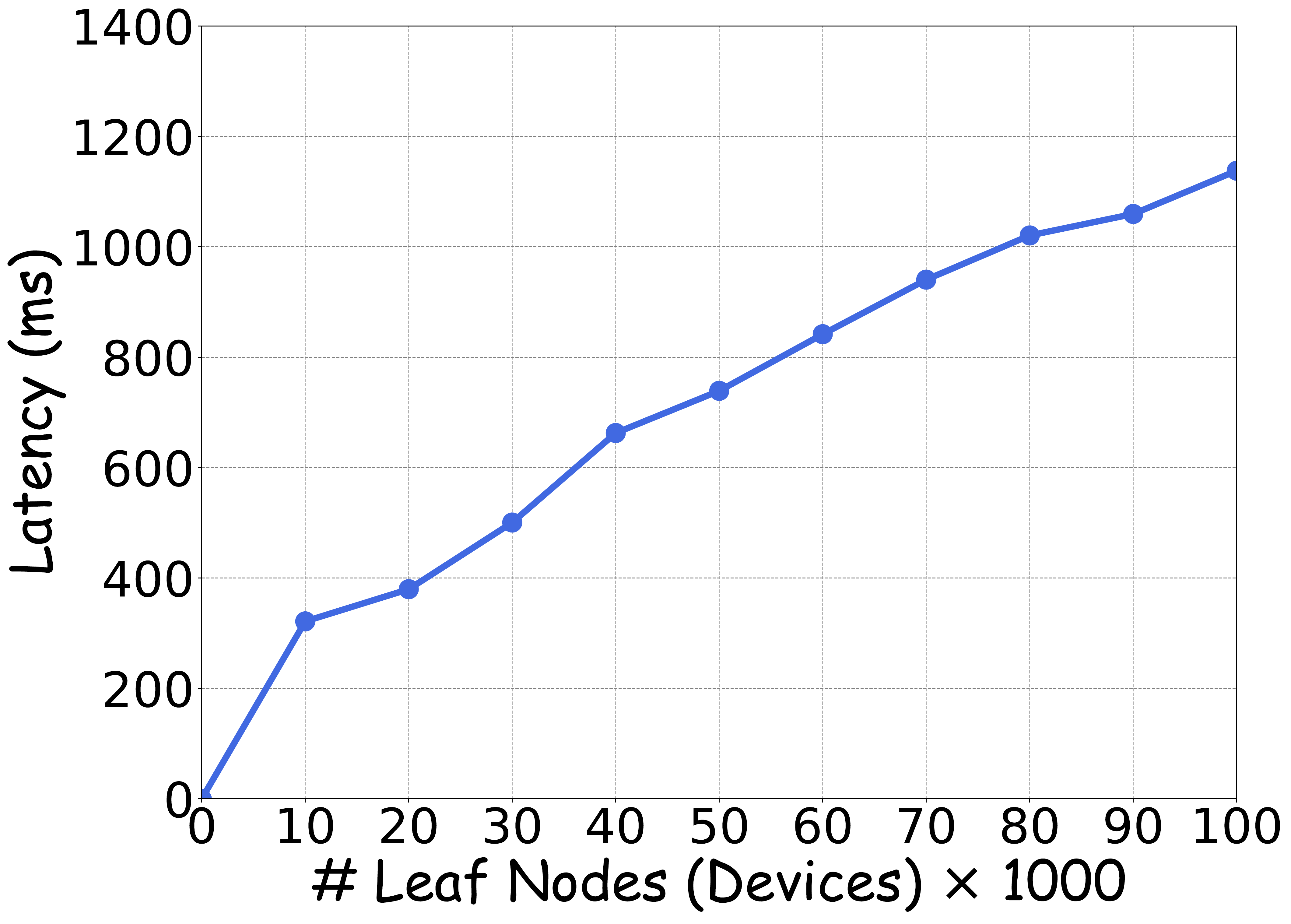}
\caption{Average latency in building abstraction trees in parallel with increasing \# leaf nodes.}\label{fig:latency_abstraction}
\end{subfigure}\hfill
\begin{subfigure}[t]{0.23\textwidth}  
\centering
\includegraphics[width=1.0\linewidth]{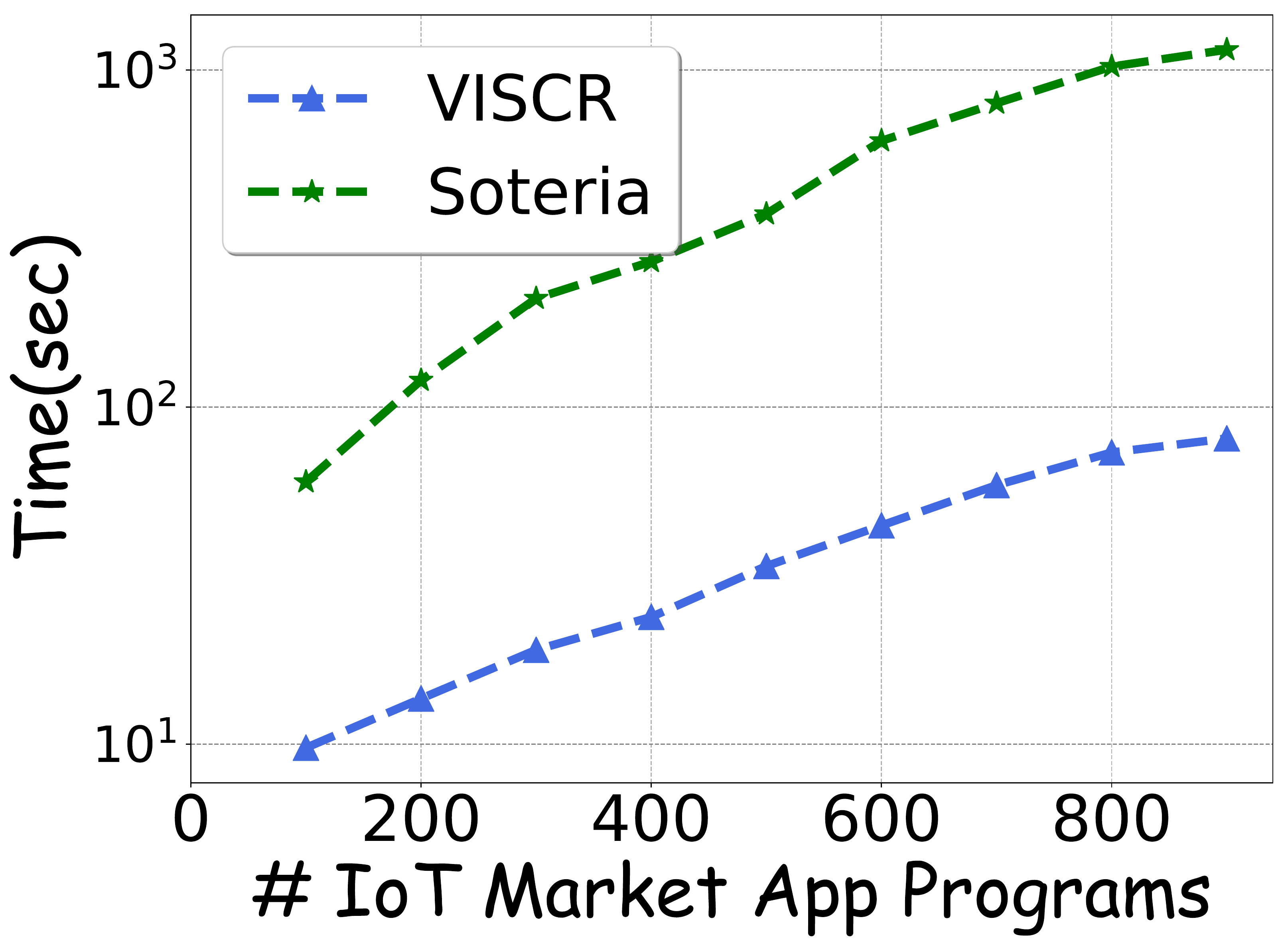}
\caption{\small Average composition latency with increasing \# IoT apps (with $\sim$30 policy attributes).}\label{fig:latency_composition}
\end{subfigure}\hfill
\begin{subfigure}[t]{0.23\textwidth} 
 \includegraphics[width=1.0\linewidth]{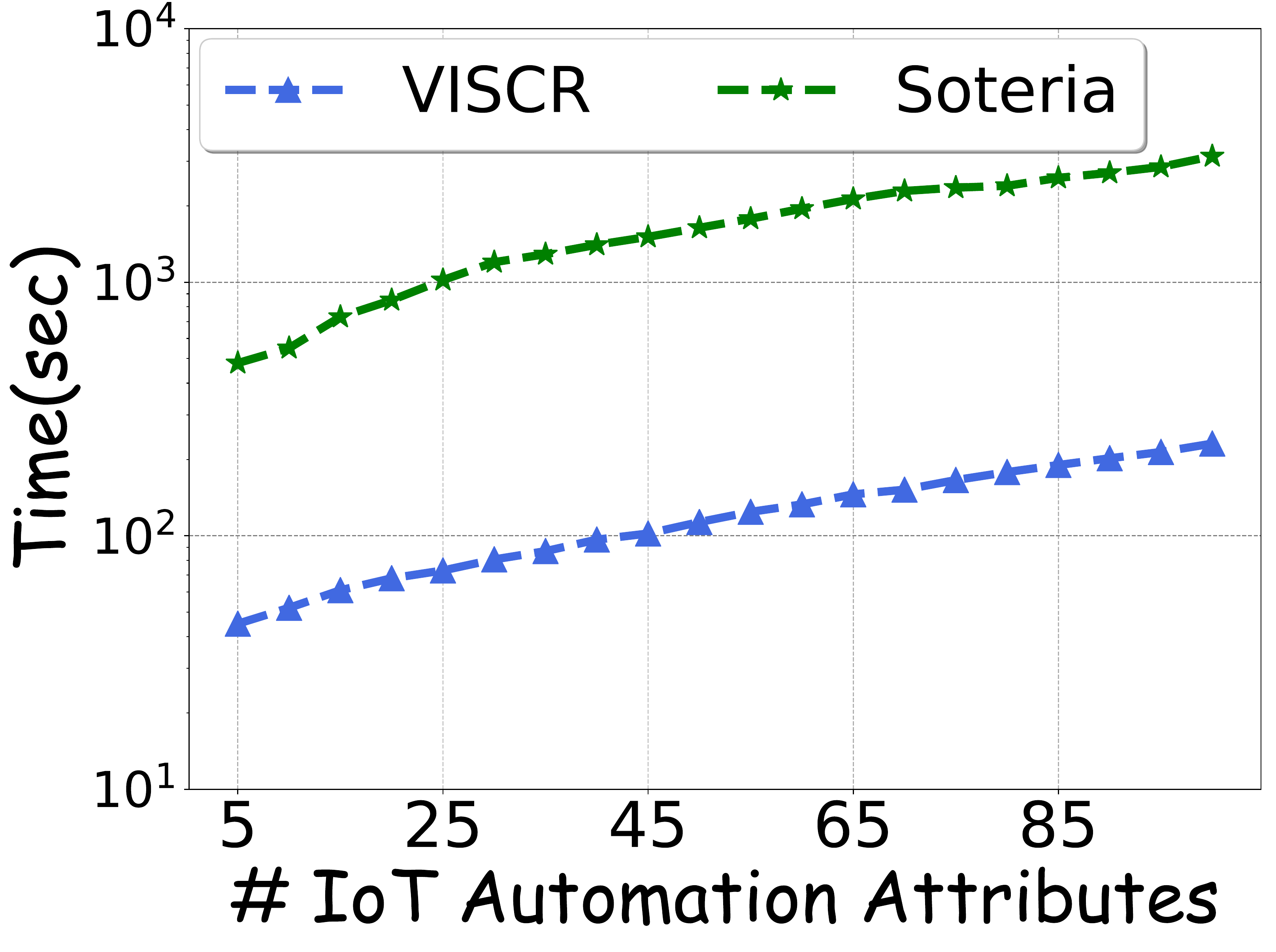}
\caption{\small Average composition latency with increasing \# policy attributes (with \numofapps IoT apps).}\label{fig:latency_composition_attributes}
\end{subfigure}\hfill
\begin{subfigure}[t]{0.23\textwidth}     
\includegraphics[width=1\linewidth]{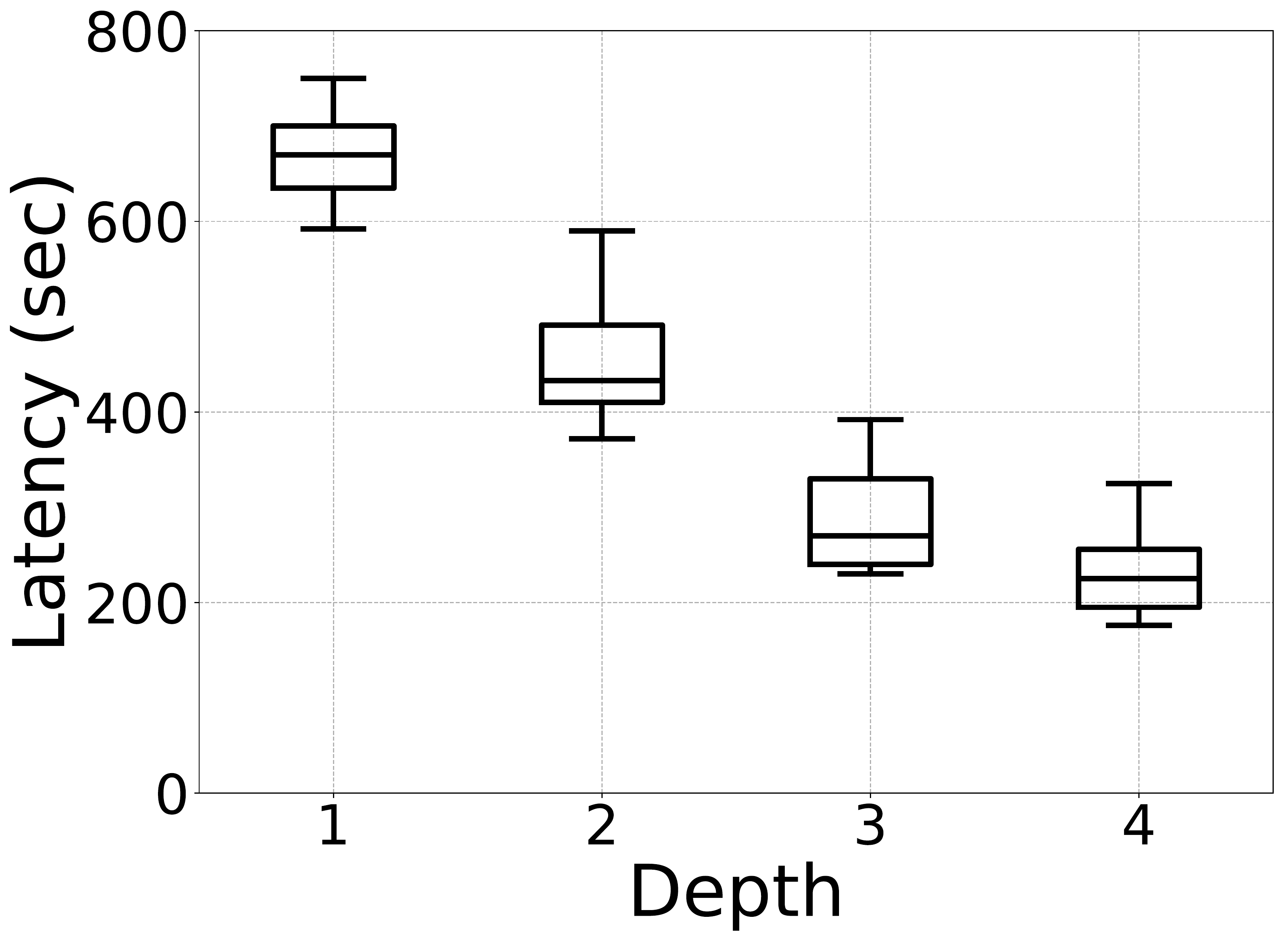}
\caption{\small Policy composition latency for $\sim$20K synthetic policies with abstraction tree depth.}\label{fig:depth_latency}
\end{subfigure}
\vspace{-0.1in}
\caption{Scalability of \secintentnospace' Policy abstraction \& Composition Engine compared to static analysis-based technique (\soterianospace).}\label{figure:policy_abstraction_composition_plots}
\vspace{-15pt}
\end{center}
\end{figure*}

We evaluate the performance of policy abstraction engine with smart city data of up to 100K devices using \texttt{DS-2}. We built upto 400 policy abstraction trees with four levels of abstractions using the dataset. For generating abstraction trees, the abstraction engine need to join hundreds of data tables performing thousands of table join operations. It look $<$1.2sec to generate upto 400 abstraction trees with 100K leaf nodes in parallel (Figure \ref{fig:latency_abstraction}). The latency performance of abstraction engine stays mostly linear considering the data join operations it performs to generate the abstraction trees. Since extracting data from the network data sources takes random times considering the network latency, we discard network latency parameter in the calculation of abstract tree generation.


\subsubsection{Security analysis} 
We evaluate the performance of our conflict detection and resolution engine with {\tt DS-1} (i.e., \numofapps IoT market apps) simulated to program the smart building IoT infrastructure (as shown in Figure \ref{figure:simulated_testbed}) and compare it with \soterianospace's violation detection mechanism. Applying model checking, we were able to find 6.6\% of the static violations across policies on {\tt DS-1}, which includes both property and state violations. As shown in Table \ref{table:policy-violations-false-positives}, \secintent on the other hand was able to find \blue{$\sim$37.7\%} of violations within same IoT apps in {\tt DS-1}. Importantly, \secintent detected 100\% of the violations captured by the static analysis technique such as \soterianospace. Among the \blue{37.7\%} of IoT apps \secintent reported 10.2\% of IoT apps that has more than one violations. Importantly, different class of violations reported by \secintent is illustrated with examples in Table \ref{table:smart-building-polcy-conflicts}. 

\begin{table}[t!]

  \footnotesize
  \centering
\begin{tabular}{|p{1.2in}|p{0.5in}|p{0.5in}|}
\hline
\Centering\bfseries Security Analysis & \Centering\bfseries \% IoT App Violations & \Centering\bfseries \% False Positives \\
  \hline

{Compile time conflicts}  & \Centering 6.6 & \Centering 0   \\ \hline
{Potential run-time violations}  & \Centering  7.9 & \Centering 1.8 \\ \hline

{Gap analysis}  & \Centering 10.4 & \Centering 1.3 \\ \hline

{Rogue Policies} & \Centering 3.8 & \Centering 0 \\ \hline

{Access violations} & \Centering 1.6 & \Centering 0 \\ \hline

{App sanity checker (SC)} & \Centering 4.2 & \Centering 0.7 \\ \hline
{Loops \& chains}  & \Centering 3.2 & \Centering 0 \\ \hline

{\textbf{Overall}}  & \Centering\bfseries 37.7 & \Centering 3.8 \\ \hline
\end{tabular}
\caption{\small Bugs and Violations detected on smart building IoT infrastructure dataset {\tt DS-1}. \% of policy violations and bugs along with the }\label{table:policy-violations-false-positives}
\vspace{-15pt}
\end{table}

In addition, \secintent identifies following major types of conflicts and violations: ($i$) gap in automation due to the inability of users to completely realize the use-case scenario resulted in 10.4\% of total IoT apps being violated with 1.3\% false positives; ($ii$) code-sanity and programming errors in temporal and spatial policies as well as policies involving specific values (e.g., temperature, humidity, time, space) resulted in 4.2\% of violations with 0.7\% false positives, where the undefined references are upto 1.1\% and unused structures are upto 2.1\% of policies; ($iii$) policies that results in access violations due prolonged access to resources beyond the specified period of time counted upto 1.6\% of IoT apps; ($iv$) rogue policies that are implemented by administrators who are not authorized to specify policies on portion of IoT infrastructures that they should not be enforcing rules (with 3.8\% of total IoT apps violated); ($v$) potential run-time violations that are detected from the policies, which includes 7.9\% of total violations; and ($vi$) finally, detected 3.2\% of loops among the automation rules that resulted in the unintended and unsafe environmental conditions. Overall, \secintent was able to detect the above discussed violations and bugs with less than 3.8\% of false positives or low intensity bugs such as Unused variables or structures and potential violations.

\subsubsection{Policy Composition} The composition cost depends on following two factors: ($i$) the number of attributes or states used and in the IoT apps, and ($ii$) the number of IoT apps. In this experiment, we perform composition with increasing number of IoT apps (i.e., in subset of 100 apps in each iteration) and capture its composition latency. We keep number of attributes constant at $\sim$30 in this experiment (i.e., the IoT apps or automation use-cases are chosen only specific to these attributes). For example, temperature is considered as an attribute, while the subcategories (e.g., high, low, different levels) of the attribute are still considered as part of the same temperature attribute. We observed that \secintent took $\sim$80.7 seconds to compose \numofapps apps, while using model checking (i.e., such as technique used in \soteria) took approximately 14.2$\times$ more time to run, which took \soteriaruntime seconds to detect the conflicts (Figure \ref{fig:latency_composition}). 

In the next experiment, we evaluate performance of conflict detection engine with increasing number of attributes and constant number of IoT apps (i.e., \numofapps). With increase in the number of attributes, the composition (i.e., conflict detection) cost of both the approaches increased. \secintent took 231 seconds to compose \numofapps IoT apps and 100 different attributes, while \soteria took $\sim$3140 seconds, \blue{which is $\sim$13$\times$ more time required to detect conflict} (Figure \ref{fig:latency_composition_attributes}). 

Following are the vital factors that contribute to improved performance (i.e., reduced conflict detection time with our graph-based composition) compared to \soterianospace: ($i$) The complex SAT formulation resulting in enumeration of all possible states required to detect the static violations. 
($ii$) Our approach optimizes the composition cost by parallelizing the translation procedure (i.e., translation of IoT apps to vendor-independent graph-based specification); ($iii$) With graph-based composition, we incrementally verify the source nodes and if the overlap exists then further into edge and target properties. This results in avoiding unnecessary comparison operation resulting in improved composition cost with our approach. and ($iv$) Finally, the graph-based composition mechanism generates the {\it composed graph islands} (i.e., policy graphs that are completely independent), when once a policy is detected as conflicting with one of the policy graph island, the policy is marked as conflicting avoiding comparison with other nodes with in the same composed graph and other graph islands. In addition, to provide fair comparison, we ran \secintent with \numofapps groovy-based apps, our composition engine took $<$92.1 seconds to compose these apps.


As \secintent also support incremental composition (i.e., to support dynamic changing IoT infrastructure), we evaluate the performance of the incremental composition as follows. We compose \numofapps IoT apps and generate the composed policy graph. We then randomly choose 100 IoT apps each time and change its attributes and allow it to incrementally recompose. 
From our experiments, it is evident that for recomposing 10 IoT app programs took $<$2.1 seconds, while recomposing 100 apps took $\sim$16.3 seconds. In normal working conditions of any IoT infrastructure, it is expected to have fewer than 10 IoT app change at any instance of time. 

Finally, we evaluate the performance of \secintent with large scale synthetic dataset ({\tt DS-3}) to emulate the smart city IoT infrastructures with $\sim$20K graph-based policies. We composed 20K graph-based policies by choosing the source and target nodes of the policies at different levels of abstraction trees i.e., choosing nodes at depth level 1 -- level 4 of abstraction tree. We run this experiment for multiple iterations to capture the average composition engine latency. For depth level = 1, the composition cost is much higher than when policy abstracts are chosen at level 4 as the nodes are chosen more towards leaf node level results in much lesser normalization cost. Therefore, 90\% of the time our tool took $<$760 seconds to compose 20K policies specified at level 1. Similarly for composing the policies specified at level 4 \secintent took $<$300 seconds.

\section{Discussion \& Limitations}\label{sec:discussion_limitations}
As part of future work, we plan to develop a more comprehensive drag-and-drop graphical interface that will allow users to express a broader range of policies. We also plan to conduct a user study that allows us to effectively evaluate the expressibility of \secintentnospace. 

\secintentnospace's tree-based abstraction engine, builds abstraction trees, which are required for specifying the graph-based automation rules or policies. The abstraction engine relies on existing IoT vendor cloud data sources and network information base (e.g., DHCP server DB, network resource DB). However, the lack of information across data sources could result in incomplete abstraction trees and policy specifications. Hence, as part of our future work, we plan to explore machine-learning-based models to fill such information gaps.

\secintent partly supports the IoT device capabilities, which are required for translating the IoT apps into graph-based specifications.  Currently, \secintent supports vendor-independent graph-based specifications for four different types of automation specification languages~\cite{groovy_smartthingsapps,openhab,decouple_ifttt,mud_spec}. We plan to support other specification mechanisms such as Apple HomeKit~\cite{apple_iot_homekit} and add more device capabilities to \secintent in our future work. Our framework provides necessary APIs that will make adding new vendor-specific automation framework much easier and seamless (as discussed in Section \ref{sec:vi_spec_code_analysis}).

Automatic inference techniques have previously been explored in the context of static code analysis for detecting bugs in  systems and web applications~\cite{engler2001bugs,pixy_static_analysis_tool,mining_temporal_specs_error_detection} and for reverse engineering protocols~\cite{prospex_kirda_proto_reverse_engg}. Similar techniques have also been used for detecting bugs and vulnerabilities in network configurations~\cite{understand_bgp_misconfig,misconfiguration_peerpressure}. We plan to explore techniques to automatically infer IoT policies from a variety of sources including the existing automation rules, IoT traffic, IoT firmware, and IoT companion apps.


\section{Related Work}\label{related}
\noindent \textbf{Intent-based Policy Frameworks.} Intent-based policies are well studied in the domain of enterprise networks ~\cite{policy_based_nw_mgmt,simplifying_nw_admin}. However, they provide limited flexibility in policy specification, while handling complex and heterogeneous IoT devices. To overcome these limitations in enterprise scenarios, recent works propose the creation high-level intent-based languages, compilers and conflict detection mechanisms ~\cite{netkat,Frenetic,practical_declarative,composing_SDN,merlin,taxonomy,maple, propane,intent_virtualization,hierarchial_policies,robotron,ravel,scenario_based,opendaylight_grouppolicy,network_intent_composition,alpaca}, and new SDN programming paradigms ~\cite{Boulder,participatory_networking,sdn_applications}. Prior efforts to develop graph-based policy specification mechanisms (PGA~\cite{pga_sigcomm15}, Janus~\cite{janus}, LMS~\cite{lms})
have focused on enterprise networks. However, these graph-based policy frameworks do not effectively handle dynamic trigger and action-based policies required by IoT devices. Hence, we propose to develop an intuitive graph-based policy framework that handles dynamic trigger and action-based policies and supports vendor-agnostic specification models to seamlessly accommodate different types of IoT programs including Groovy, OpenHAB, IFTTT, and MUD profiles.

\vspace{3pt}
\noindent \textbf{Conflict-Detection \& Verification with Dynamic IoT Policies.} Verification and testing of dynamic policies for middleboxes are well-studied problems~\cite{snap,flowtest,buzz,flowtags,sfc_checker,automatic_synthesis,sla_verifier,improving_nw_mgmt}. Unlike our work, prior studies do not address the dynamic group-based policy requirement of IoT infrastructures to handle safety, security, and privacy policies. Recent efforts have attempted to use formal verification techniques and static taint tracking to verify the correctness of deployed automation policies in homogeneous IoT environments~\cite{soteria,iotmon,iot2_verify, iot_policy_confict_planning,saint_atc}. Similarly, recently proposed works highlighted the need for novel access control models and policies to secure IoT infrastructures~\cite{Schuster:situational_acl_ccs2018,rethink_acls_iot}. However, existing IoT infrastructures are dynamic with diverse IoT devices, programmed using heterogeneous programming frameworks, that make static-verification techniques ineffective. Also, a few recent studies that focus on identifying the policy conflicts arising  in complex smart-city infrastructures~\cite{cityguard,watchdog} do not deal with security or privacy issues. Similarly, Soteria~\cite{soteria}, developed an intermediate representation for Groovy-based IoT polices and used model checking on it to verify the properties. We propose to build a vendor-independent model, that allows automation rules or policies, specified using multiple commodity IoT apps to be translated into vendor-independent policy-specification graphs for robust and proactive conflict detection and resolution.

\section{Conclusion}\label{sec:conclusion}
Emerging consumer IoT infrastructures are characterized by a growing number of heterogeneous devices. \secintent provides a unified policy engine that allows for conflict-free policy specification and enforcement in IoT infrastructure. \secintent achieves this by unifying policy abstractions, automatically extracting IoT infrastructure topology and converting diverse policy languages such as Groovy-based SmartThings, OpenHAB, IFTTT-based templates, and MUD-based profiles into a vendor-independent graph-based specification. These abstractions enable \secintent to detect rouge policies, bugs, and conflicts. They also allow for easier specification and efficient composition of dynamic policy intents, from users and administrators. 
In a dataset of \numofapps IoT market apps with a mix of Groovy, OpeHAB, IFTTT, and MUD-based policies, \secintent detected conflicts in 342 apps and provided resolution mechanism to it in under 81 seconds and can adapt new policies with sub-second latency.

\bibliographystyle{unsrt}
\bibliography{ref2}

\begin{thebibliography}{10}

\bibitem{gartner-future-smart-home}
{The Future Smart Home: 500 Smart Objects Will Enable New Business
  Opportunities.}, March 2014.
\newblock \url{http://www.gartner.com/newsroom/id/2839717}.

\bibitem{gartner_2020}
{Gartner Says 8.4 Billion Connected "Things" Will Be in Use in 2017, Up 31
  Percent From 2016}, November 2017.
\newblock \url{https://www.gartner.com/newsroom/id/3598917}.

\bibitem{openhab}
OpenHAB~Textual Rules, October 2018.
\newblock \url{https://www.openhab.org/docs/configuration/rules-dsl.html}.

\bibitem{apple_iot_homekit}
Apple's~Homekit. Accessed.
\newblock March 2017.
\newblock \url{https://developer.apple.com/homekit/}.

\bibitem{groovy_smartthingsapps}
{Samsung SmartThings Public GitHub Repo. Accessed}.
\newblock 2017.
\newblock
  \url{https://github.com/SmartThingsCommunity/SmartThingsPublic/blob/master/smartapps/smartthings/camera-power-scheduler.src/camera-power-scheduler.groovy}.

\bibitem{mud_spec}
Manufacturer Usage Description~(MUD) Specification, October 2018.
\newblock \url{https://tools.ietf.org/html/draft-ietf-opsawg-mud-25}.

\bibitem{decouple_ifttt}
Earlence Fernandes, Amir Rahmati, Jaeyeon Jung, and Atul Prakash.
\newblock Decoupled-ifttt: Constraining privilege in trigger-action platforms
  for the internet of things.
\newblock {\em CoRR}, abs/1707.00405, 2017.

\bibitem{rethink_acls_iot}
Weijia He, Maximilian Golla, Roshni Padhi, Jordan Ofek, Markus D{\"u}rmuth,
  Earlence Fernandes, and Blase Ur.
\newblock Rethinking access control and authentication for the home internet of
  things (iot).
\newblock In {\em 27th {USENIX} Security Symposium ({USENIX} Security 18)},
  pages 255--272, Baltimore, MD, 2018. {USENIX} Association.

\bibitem{soteria}
Z.~Berkay Celik, Patrick McDaniel, and Gang Tan.
\newblock Soteria: Automated iot safety and security analysis.
\newblock In {\em 2018 {USENIX} Annual Technical Conference ({USENIX} {ATC}
  18)}, pages 147--158, Boston, MA, 2018. {USENIX} Association.

\bibitem{celikiotguard}
Z~Berkay Celik, Gang Tan, and Patrick McDaniel.
\newblock Iotguard: Dynamic enforcement of security and safety policy in
  commodity iot.

\bibitem{apple_iot_homekit_swift}
HomeKit Tutorial:~Getting Started.
\newblock July 2018.
\newblock
  \url{https://www.raywenderlich.com/5313-homekit-tutorial-getting-started}.

\bibitem{openhab_missing_quotes}
OpenHAB:~Rules not running.
\newblock May 2018.
\newblock \url{https://community.openhab.org/t/solved-rules-not-running/45776}.

\bibitem{openhab_missing_if_braces}
OpenHAB:~Turn light OFF based~on timer.
\newblock March 2017.
\newblock
  \url{https://community.openhab.org/t/turn-light-off-based-on-timer/9558/30}.

\bibitem{openhab_misbehaving_val_var_references}
OpenHAB:~File level val/var declarations~behave unexpectedly.
\newblock March 2017.
\newblock
  \url{https://community.openhab.org/t/file-level-val-var-declarations-behave-unexpectedly/25304/3}.

\bibitem{smartthings_errors_community}
SmartThings~Community Discussions.
\newblock March 2019.
\newblock \url{https://community.smartthings.com/c/smartapps}.

\bibitem{ifttt_errors_community}
Garadget~IFTTT Errors.
\newblock March 2019.
\newblock \url{https://community.garadget.com/t/ifttt-errors/4103}.

\bibitem{applehomekit_errors_community}
Apple~HomePod Communities.
\newblock March 2019.
\newblock \url{https://discussions.apple.com/community/homepod}.

\bibitem{samsung_ifttt}
{IFTTT: Samsung SmartThings. Accessed}.
\newblock 2017.
\newblock \url{https://ifttt.com/smartthings}.

\bibitem{iot_vulnerability1}
{The 5 Worst Examples of IoT Hacking and Vulnerabilities in Recorded History},
  May 2017.
\newblock \url{
  https://www.iotforall.com/5-worst-iot-hacking-vulnerabilities/}.

\bibitem{iot_vulnerability2}
{ISTR, Internet Security Threat Report}, April 2016.
\newblock \url{
  https://www.symantec.com/content/dam/symantec/docs/reports/istr-21-2016-en.pdf}.

\bibitem{pga_sigcomm15}
Chaithan Prakash, Jeongkeun Lee, Yoshio Turner, Joon-Myung Kang, Aditya Akella,
  Sujata Banerjee, Charles Clark, Yadi Ma, Puneet Sharma, and Ying Zhang.
\newblock Pga: Using graphs to express and automatically reconcile network
  policies.
\newblock In {\em Proceedings of the 2015 ACM Conference on Special Interest
  Group on Data Communication}, SIGCOMM '15, pages 29--42, New York, NY, USA,
  2015. ACM.

\bibitem{nest_iot_cloud_apis}
{Nest Cloud APIs}.
\newblock April 2017.
\newblock \url{https://developers.nest.com/documentation/cloud/get-started}.

\bibitem{samsung_iot_cloud_apis}
{Samsung Cloud APIs}.
\newblock April 2017.
\newblock \url{http://developer.samsung.com/smart-home}.

\bibitem{multi_label_binarizer}
{Sklearn: LabelBinarizer}.
\newblock
  \url{https://scikit-learn.org/stable/modules/generated/sklearn.preprocessing.LabelBinarizer.html}.

\bibitem{k-means_clustering_states}
{Using the elbow method to determine the optimal number of clusters for k-means
  clustering }.
\newblock \url{https://bl.ocks.org/rpgove/0060ff3b656618e9136b}.

\bibitem{z-ranking_statistical_engler}
Ted Kremenek and Dawson Engler.
\newblock Z-ranking: Using statistical analysis to counter the impact of static
  analysis approximations.
\newblock In {\em International Static Analysis Symposium}, pages 295--315.
  Springer, 2003.

\bibitem{correlation_ranking_engler}
Ted Kremenek, Ken Ashcraft, Junfeng Yang, and Dawson Engler.
\newblock Correlation exploitation in error ranking.
\newblock In {\em ACM SIGSOFT Software Engineering Notes}, volume~29, pages
  83--93. ACM, 2004.

\bibitem{jen_rule_placement}
Nanxi Kang, Zhenming Liu, Jennifer Rexford, and David Walker.
\newblock Optimizing the "one big switch" abstraction in software-defined
  networks.
\newblock In {\em Proceedings of the Ninth ACM Conference on Emerging
  Networking Experiments and Technologies}, CoNEXT '13, pages 13--24, New York,
  NY, USA, 2013. ACM.

\bibitem{adoptable_rule_placement}
S.~{Zhang}, F.~{Ivancic}, C.~{Lumezanu}, Y.~{Yuan}, A.~{Gupta}, and S.~{Malik}.
\newblock An adaptable rule placement for software-defined networks.
\newblock In {\em 2014 44th Annual IEEE/IFIP International Conference on
  Dependable Systems and Networks}, pages 88--99, June 2014.

\bibitem{congress_arch}
{\em {Congress Architecture}, 2018}.
\newblock \url{http://congress.readthedocs.io/en/latest/architecture.html}.

\bibitem{networkx_nxgraph}
{Networkx Graph Creation. Accessed}.
\newblock December 2017.
\newblock
  \url{https://networkx.github.io/documentation/networkx-1.7/tutorial/tutorial.html}.

\bibitem{graphviz}
{\em {Graphviz - Graph Visualization Software}}.
\newblock \url{https://www.graphviz.org/}.

\bibitem{iot_test_bench}
{IoT TestBench: A micro-benchmark suite to assess the effectiveness of tools
  designed for IoT apps }.
\newblock \url{https://github.com/IoTBench/IoTBench-test-suite}.

\bibitem{smart-home-data}
Sean Barker, Aditya Mishra, David Irwin, Emmanuel Cecchet, Prashant Shenoy, and
  Jeannie Albrecht.
\newblock Smart*: An open data set and tools for enabling research in
  sustainable homes.
\newblock {\em SustKDD, August}, 111:112, 2012.

\bibitem{smart_homedata1}
{Smart Home Data Set for Sustainability)}.
\newblock December 2017.
\newblock \url{http://traces.cs.umass.edu/index.php/Smart/Smart}.

\bibitem{sfo_dataset_all}
{SFO City scale data set (City facilities)}.
\newblock December 2017.
\newblock \url{https://data.sfgov.org}.

\bibitem{smart-city-survey}
Charith Perera, Yongrui Qin, Julio~C. Estrella, Stephan Reiff-Marganiec, and
  Athanasios~V. Vasilakos.
\newblock Fog computing for sustainable smart cities: A survey.
\newblock {\em ACM Comput. Surv.}, 50(3):32:1--32:43, June 2017.

\bibitem{viscr_secintent_dataset}
{Secure, Safe and Privacy With IoT Infrastructures Policy Framework and
  Dataset. Accessed}.
\newblock March 2019.
\newblock {Anonymized for double blinded review.}

\bibitem{sfo_dataset1}
{SFO City scale data set (Locations and Boundaries)}.
\newblock December 2017.
\newblock
  \url{https://data.sfgov.org/Geographic-Locations-and-Boundaries/List-of-Streets-and-Intersections/pu5n-qu5c}.

\bibitem{sfo_dataset2}
{SFO City scale data set (Traffic Signals)}.
\newblock December 2017.
\newblock
  \url{https://data.sfgov.org/Transportation/Map-of-Traffic-Signals/8xta-sna8}.

\bibitem{sfo_dataset3}
{SFO City scale data set (City facilities)}.
\newblock December 2017.
\newblock
  \url{https://data.sfgov.org/City-Infrastructure/Map-of-City-Facilities/bps8-63cu}.

\bibitem{engler2001bugs}
Dawson Engler, David~Yu Chen, Seth Hallem, Andy Chou, and Benjamin Chelf.
\newblock Bugs as deviant behavior: A general approach to inferring errors in
  systems code.
\newblock In {\em ACM SIGOPS Operating Systems Review}, volume~35, pages
  57--72. ACM, 2001.

\bibitem{pixy_static_analysis_tool}
E.~Kirda, N.~Jovanovic, and C.~Kruegel.
\newblock {Pixy: A Static Analysis Tool for Detecting Web Application
  Vulnerabilities (Short Paper)}.
\newblock In {\em {2006 IEEE Symposium on Security and Privacy (S\&P'06)(SP)}},
  volume~00, pages 258--263, 05 2006.

\bibitem{mining_temporal_specs_error_detection}
Westley Weimer and George~C. Necula.
\newblock Mining temporal specifications for error detection.
\newblock In Nicolas Halbwachs and Lenore~D. Zuck, editors, {\em Tools and
  Algorithms for the Construction and Analysis of Systems}, pages 461--476,
  Berlin, Heidelberg, 2005. Springer Berlin Heidelberg.

\bibitem{prospex_kirda_proto_reverse_engg}
Paolo~Milani Comparetti, Gilbert Wondracek, Christopher Kruegel, and Engin
  Kirda.
\newblock Prospex: Protocol specification extraction.
\newblock In {\em Security and Privacy, 2009 30th IEEE Symposium on}, pages
  110--125. IEEE, 2009.

\bibitem{understand_bgp_misconfig}
Ratul Mahajan, David Wetherall, and Tom Anderson.
\newblock Understanding bgp misconfiguration.
\newblock {\em SIGCOMM Comput. Commun. Rev.}, 32(4):3--16, August 2002.

\bibitem{misconfiguration_peerpressure}
Helen~J. Wang, John~C. Platt, Yu~Chen, Ruyun Zhang, and Yi-Min Wang.
\newblock Automatic misconfiguration troubleshooting with peerpressure.
\newblock In {\em Proceedings of the 6th Conference on Symposium on Opearting
  Systems Design \& Implementation - Volume 6}, OSDI'04, pages 17--17,
  Berkeley, CA, USA, 2004. USENIX Association.

\bibitem{policy_based_nw_mgmt}
John Strassner.
\newblock {\em Policy-Based Network Management: Solutions for the Next
  Generation (The Morgan Kaufmann Series in Networking)}.
\newblock Morgan Kaufmann Publishers Inc., San Francisco, CA, USA, 2003.

\bibitem{simplifying_nw_admin}
D.~C. Verma.
\newblock {Simplifying network administration using policy-based management}.
\newblock {\em IEEE Network}, 16(2):20--26, Mar 2002.

\bibitem{netkat}
Steffen Smolka, Spiridon Eliopoulos, Nate Foster, and Arjun Guha.
\newblock A fast compiler for netkat.
\newblock In {\em Proceedings of the 20th ACM SIGPLAN International Conference
  on Functional Programming}, ICFP 2015, pages 328--341, New York, NY, USA,
  2015. ACM.

\bibitem{Frenetic}
Nate Foster, Rob Harrison, Michael~J. Freedman, Christopher Monsanto, Jennifer
  Rexford, Alec Story, and David Walker.
\newblock Frenetic: A network programming language.
\newblock In {\em Proceedings of the 16th ACM SIGPLAN International Conference
  on Functional Programming}, ICFP '11, pages 279--291, New York, NY, USA,
  2011. ACM.

\bibitem{practical_declarative}
Timothy~L. Hinrichs, Natasha~S. Gude, Martin Casado, John~C. Mitchell, and
  Scott Shenker.
\newblock Practical declarative network management.
\newblock In {\em Proceedings of the 1st ACM Workshop on Research on Enterprise
  Networking}, WREN '09, pages 1--10, New York, NY, USA, 2009. ACM.

\bibitem{composing_SDN}
Christopher Monsanto, Joshua Reich, Nate Foster, Jennifer Rexford, and David
  Walker.
\newblock Composing software-defined networks.
\newblock In {\em Proceedings of the 10th USENIX Conference on Networked
  Systems Design and Implementation}, nsdi'13, pages 1--14, Berkeley, CA, USA,
  2013. USENIX Association.

\bibitem{merlin}
Robert Soul{\'e}, Shrutarshi Basu, Parisa~Jalili Marandi, Fernando Pedone,
  Robert Kleinberg, Emin~Gun Sirer, and Nate Foster.
\newblock Merlin: A language for provisioning network resources.
\newblock In {\em Proceedings of the 10th ACM International on Conference on
  Emerging Networking Experiments and Technologies}, CoNEXT '14, pages
  213--226, New York, NY, USA, 2014. ACM.

\bibitem{taxonomy}
C.~Trois, M.~D.~Del Fabro, L.~C.~E. de~Bona, and M.~Martinello.
\newblock A survey on sdn programming languages: Toward a taxonomy.
\newblock {\em IEEE Communications Surveys Tutorials}, 18(4):2687--2712,
  Fourthquarter 2016.

\bibitem{maple}
Andreas Voellmy, Junchang Wang, Y~Richard Yang, Bryan Ford, and Paul Hudak.
\newblock Maple: Simplifying sdn programming using algorithmic policies.
\newblock In {\em Proceedings of the ACM SIGCOMM 2013 Conference on SIGCOMM},
  SIGCOMM '13, pages 87--98, New York, NY, USA, 2013. ACM.

\bibitem{propane}
Ryan Beckett, Ratul Mahajan, Todd Millstein, Jitendra Padhye, and David Walker.
\newblock Don't mind the gap: Bridging network-wide objectives and device-level
  configurations.
\newblock In {\em Proceedings of the 2016 Conference on ACM SIGCOMM 2016
  Conference}, SIGCOMM '16, pages 328--341, New York, NY, USA, 2016. ACM.

\bibitem{intent_virtualization}
R.~Cohen, K.~Barabash, B.~Rochwerger, L.~Schour, D.~Crisan, R.~Birke,
  C.~Minkenberg, M.~Gusat, R.~Recio, and V.~Jain.
\newblock An intent-based approach for network virtualization.
\newblock In {\em 2013 IFIP/IEEE International Symposium on Integrated Network
  Management (IM 2013)}, pages 42--50, May 2013.

\bibitem{hierarchial_policies}
Andrew~D. Ferguson, Arjun Guha, Chen Liang, Rodrigo Fonseca, and Shriram
  Krishnamurthi.
\newblock Hierarchical policies for software defined networks.
\newblock In {\em Proceedings of the First Workshop on Hot Topics in Software
  Defined Networks}, HotSDN '12, pages 37--42, New York, NY, USA, 2012. ACM.

\bibitem{robotron}
Yu-Wei~Eric Sung, Xiaozheng Tie, Starsky~H.Y. Wong, and Hongyi Zeng.
\newblock {Robotron: Top-down Network Management at Facebook Scale}.
\newblock In {\em Proceedings of the 2016 ACM SIGCOMM Conference}, SIGCOMM '16,
  pages 426--439, New York, NY, USA, 2016. ACM.

\bibitem{ravel}
Anduo Wang, Xueyuan Mei, Jason Croft, Matthew Caesar, and Brighten Godfrey.
\newblock {Ravel: A Database-Defined Network}.
\newblock In {\em Proceedings of the Symposium on SDN Research}, SOSR '16,
  pages 5:1--5:7, New York, NY, USA, 2016. ACM.

\bibitem{scenario_based}
Yifei Yuan, Dong Lin, Rajeev Alur, and Boon~Thau Loo.
\newblock Scenario-based programming for sdn policies.
\newblock In {\em Proceedings of the 11th ACM Conference on Emerging Networking
  Experiments and Technologies}, CoNEXT '15, pages 34:1--34:13, New York, NY,
  USA, 2015. ACM.

\bibitem{opendaylight_grouppolicy}
OpenDaylight Group~Policy. Accessed.
\newblock 2017.
\newblock \url{https://wiki.opendaylight.org/view/Group_Policy:Main}.

\bibitem{network_intent_composition}
{Anu Mercian, Felipe Yrineu, Joon-Myung Kang, Raphael Amorim, Saket M Mahajani,
  Mario Sanchez and Sujata Banerjee}.
\newblock {\em Network Intent Composition (NIC) Be Feature Update and Demo:
  Intent Compilation, Lifecycle Management and Automated Mapping}, Presented in
  OpenDaylight Summit 2016.
\newblock \url{http://sched.co/7RBY}.

\bibitem{alpaca}
Nanxi Kang, Ori Rottenstreich, Sanjay Rao, and Jennifer Rexford.
\newblock Alpaca: Compact network policies with attribute-carrying addresses.
\newblock In {\em Proceedings of the 11th ACM Conference on Emerging Networking
  Experiments and Technologies}, CoNEXT '15, pages 7:1--7:13, New York, NY,
  USA, 2015. ACM.

\bibitem{Boulder}
Boulder:~Intent based North Bound Interface~({NBI}).
\newblock February 2015.
\newblock
  \url{https://www.opennetworking.org/images/stories/sdn-solution-showcase/germany2015/Boulder%20-%20Intent%20Based%20NBI.pdf}.

\bibitem{participatory_networking}
Andrew~D. Ferguson, Arjun Guha, Chen Liang, Rodrigo Fonseca, and Shriram
  Krishnamurthi.
\newblock Participatory networking: An api for application control of sdns.
\newblock In {\em Proceedings of the ACM SIGCOMM 2013 Conference on SIGCOMM},
  SIGCOMM '13, pages 327--338, New York, NY, USA, 2013. ACM.

\bibitem{sdn_applications}
M.~Pham and D.~B. Hoang.
\newblock Sdn applications - the intent-based northbound interface realisation
  for extended applications.
\newblock In {\em 2016 IEEE NetSoft Conference and Workshops (NetSoft)}, pages
  372--377, June 2016.

\bibitem{janus}
Anubhavnidhi Abhashkumar, Joon-Myung Kang, Sujata Banerjee, Aditya Akella, Ying
  Zhang, and Wenfei Wu.
\newblock {Supporting Diverse Dynamic Intent-based Policies Using Janus}.
\newblock In {\em Proceedings of the 13th International Conference on Emerging
  Networking EXperiments and Technologies}, CoNEXT '17, pages 296--309, New
  York, NY, USA, 2017. ACM.

\bibitem{lms}
J.~M. Kang, J.~Lee, V.~Nagendra, and S.~Banerjee.
\newblock {LMS: Label Management Service for intent-driven Cloud Management}.
\newblock In {\em 2017 IFIP/IEEE Symposium on Integrated Network and Service
  Management (IM)}, pages 177--185, May 2017.

\bibitem{snap}
Mina~Tahmasbi Arashloo, Yaron Koral, Michael Greenberg, Jennifer Rexford, and
  David Walker.
\newblock {SNAP:} stateful network-wide abstractions for packet processing.
\newblock {\em CoRR}, abs/1512.00822, 2015.

\bibitem{flowtest}
Seyed~K. Fayaz and Vyas Sekar.
\newblock Testing stateful and dynamic data planes with flowtest.
\newblock In {\em Proceedings of the Third Workshop on Hot Topics in Software
  Defined Networking}, HotSDN '14, pages 79--84, New York, NY, USA, 2014. ACM.

\bibitem{buzz}
Seyed~K. Fayaz, Tianlong Yu, Yoshiaki Tobioka, Sagar Chaki, and Vyas Sekar.
\newblock Buzz: Testing context-dependent policies in stateful networks.
\newblock In {\em Proceedings of the 13th Usenix Conference on Networked
  Systems Design and Implementation}, NSDI'16, pages 275--289, Berkeley, CA,
  USA, 2016. USENIX Association.

\bibitem{flowtags}
Seyed~Kaveh Fayazbakhsh, Vyas Sekar, Minlan Yu, and Jeffrey~C. Mogul.
\newblock Flowtags: Enforcing network-wide policies in the presence of dynamic
  middlebox actions.
\newblock In {\em Proceedings of the Second ACM SIGCOMM Workshop on Hot Topics
  in Software Defined Networking}, HotSDN '13, pages 19--24, New York, NY, USA,
  2013. ACM.

\bibitem{sfc_checker}
B.~Tschaen, Y.~Zhang, T.~Benson, S.~Banerjee, J.~Lee, and J.~M. Kang.
\newblock Sfc-checker: Checking the correct forwarding behavior of service
  function chaining.
\newblock In {\em 2016 IEEE Conference on Network Function Virtualization and
  Software Defined Networks (NFV-SDN)}, pages 134--140, Nov 2016.

\bibitem{automatic_synthesis}
Wenfei Wu, Ying Zhang, and Sujata Banerjee.
\newblock Automatic synthesis of nf models by program analysis.
\newblock In {\em Proceedings of the 15th ACM Workshop on Hot Topics in
  Networks}, HotNets '16, pages 29--35, New York, NY, USA, 2016. ACM.

\bibitem{sla_verifier}
Y.~Zhang, W.~Wu, S.~Banerjee, J.~M. Kang, and M.~A. Sanchez.
\newblock Sla-verifier: Stateful and quantitative verification for service
  chaining.
\newblock In {\em IEEE INFOCOM 2017 - IEEE Conference on Computer
  Communications}, pages 1--9, May 2017.

\bibitem{improving_nw_mgmt}
H.~Kim and N.~Feamster.
\newblock Improving network management with software defined networking.
\newblock {\em IEEE Communications Magazine}, 51(2):114--119, February 2013.

\bibitem{iotmon}
Wenbo Ding and Hongxin Hu.
\newblock On the safety of iot device physical interaction control.
\newblock In {\em Proceedings of the 2018 ACM SIGSAC Conference on Computer and
  Communications Security}, CCS '18, pages 832--846, New York, NY, USA, 2018.
  ACM.

\bibitem{iot2_verify}
Abdullah~Al Farooq, Ehab Al{-}Shaer, Thomas Moyer, and Krishna Kant.
\newblock Iotc2: {A} formal method approach for detecting conflicts in large
  scale iot systems.
\newblock {\em CoRR}, abs/1812.03966, 2018.

\bibitem{iot_policy_confict_planning}
Emre Göynügür, Sara Bernardini, Geeth Mel, Kartik Talamadupula, and Murat
  Sensoy.
\newblock Policy conflict resolution in iot via planning.
\newblock pages 169--175, 04 2017.

\bibitem{saint_atc}
Z.~Berkay Celik, Leonardo Babun, Amit~Kumar Sikder, Hidayet Aksu, Gang Tan,
  Patrick McDaniel, and A.~Selcuk Uluagac.
\newblock Sensitive information tracking in commodity iot.
\newblock In {\em 27th {USENIX} Security Symposium ({USENIX} Security 18)},
  pages 1687--1704, Baltimore, MD, 2018. {USENIX} Association.

\bibitem{Schuster:situational_acl_ccs2018}
Roei Schuster, Vitaly Shmatikov, and Eran Tromer.
\newblock Situational access control in the internet of things.
\newblock In {\em Proceedings of the 2018 ACM SIGSAC Conference on Computer and
  Communications Security}, CCS '18, pages 1056--1073, New York, NY, USA, 2018.
  ACM.

\bibitem{cityguard}
Meiyi Ma, Sarah~Masud Preum, and John~A. Stankovic.
\newblock {CityGuard: A Watchdog for Safety-Aware Conflict Detection in Smart
  Cities}.
\newblock In {\em Proceedings of the Second International Conference on
  Internet-of-Things Design and Implementation}, IoTDI '17, pages 259--270, New
  York, NY, USA, 2017. ACM.

\bibitem{watchdog}
M.~Ma, S.~M. Preum, W.~Tarneberg, M.~Ahmed, M.~Ruiters, and J.~Stankovic.
\newblock {Detection of Runtime Conflicts among Services in Smart Cities}.
\newblock In {\em 2016 IEEE International Conference on Smart Computing
  (SMARTCOMP)}, pages 1--10, May 2016.

\end{thebibliography}

\clearpage
\appendix







\end{document}